\documentclass[preprint,preprintnumbers,amsmath,amssymb,nofootinbib,secnumarabic,superscriptaddress]{revtex4}
\usepackage{amsfonts}
\usepackage{amsmath}
\usepackage{graphicx,psfrag,natbib}
\usepackage{subfig}
\numberwithin{equation}{section}

\def\be{\begin{equation} }
\def\ee{\end{equation} }
\def\la{\langle}
\def\ra{\rangle}

\begin{document}
\title{On graviton non-gaussianities during inflation}

\author{Juan M. Maldacena}
\email{malda@sns.ias.edu}
\affiliation{School of Natural Sciences, Institute for Advanced Study, Princeton, NJ, 08540}
\author{Guilherme L. Pimentel}
 \email{gpimente@princeton.edu}
\affiliation{
Joseph Henry Laboratories, Princeton University, Princeton, NJ, 08544.
}

\preprint{PUPT-2371}

\begin{abstract}

We consider the most general three point function for gravitational waves produced during a
period of exactly de Sitter expansion.  The de Sitter isometries constrain the possible shapes to
only three: two preserving parity and one violating parity.  These isometries
imply that these correlation functions should be conformal invariant.
One of the shapes is produced by the ordinary gravity action.  The other shape is produced by
a higher derivative correction and could  be as large as the gravity contribution.
The parity violating
shape does not contribute to the bispectrum  \cite{Soda:2011am,Shiraishi:2011st}, even though it
is present in the wavefunction.
We also introduce a spinor helicity formalism to describe de Sitter
gravitational waves with circular polarization.

These results also apply to correlation functions in Anti-de Sitter space. They
also describe the general form
of stress tensor correlation functions, in momentum space,  in a three dimensional conformal field theory.
Here all three shapes can arise, including the parity violating one.

\end{abstract}
\maketitle
\pagebreak
\section{Introduction}

Recently there has been some effort in understanding the non-gaussian corrections to
primordial fluctuations generated during inflation. The simplest correction is a contribution to
the three point functions of scalar and tensor fluctuations.
For scalar fluctuations there is a classification of the possible shapes for the three point function
that appear to the leading orders in the derivative expansion for the scalar field \cite{Cheung:2007st,Cheung:2007sv,Chen:2006nt,Weinberg:2008hq}.

In this paper we consider tensor fluctuations. We work in the de Sitter approximation and we argue
that there are only three possible shapes for the three point function {\it to all orders} in the
derivative expansion. Thus the de Sitter approximation allows us to consider arbitrarily high order
corrections in the derivative expansion. The idea is simply that the three point function is
constrained by the de Sitter isometries. At late times, the interesting part of the  wavefunction becomes
 time independent and the de Sitter isometries act as the conformal group on the spatial boundary.
 We are familiar with the exact  scale invariance, but,
 in addition, we also have  conformal invariance. The conformal invariance fixes the three point functions almost uniquely.
By ``almost", we simply mean that there are three possible shapes allowed, two that preserve parity and
one that violates parity. We compute explicitly these shapes and we show that they are the only
ones consistent with the conformal symmetry. In particular, we analyze in detail the
 constraints from   conformal invariance.
In order to compute these three shapes it is enough to compute them for a simple Lagrangian that
is general enough to produce them. The Einstein gravity Lagrangian produces one of these three shapes
 \cite{Arutyunov:1999nw,Maldacena:2002vr}. The other parity conserving shape can be obtained
by adding a $\int W^3$ term to the action, where $W$ is the Weyl tensor. Finally,  the parity
violating shape can be obtained by adding $\int W^2 \widetilde W$, where $\widetilde W$ is the Weyl tensor with two indices
contracted with an $\epsilon$ tensor. The fact that the gravitational wave expectation value is determined by the
symmetries is intimately connected with the following fact: in four dimensional flat space there are
also three possible three point graviton
scattering amplitudes \cite{Benincasa:2007xk} \footnote{ In flat space, one
has to complexify the momenta to
have  nonvanishing three point amplitudes.  In de Sitter, they are the natural observables.}.
Though the parity violating shape is contained in the wavefunction of the universe (or in related AdS
partition functions), it does not arise for expectation values \cite{Soda:2011am,Shiraishi:2011st}.\footnote{
The previous version of this paper incorrectly stated that the parity violating shape contributed
to de Sitter expectation values. The fact that there is no parity violation in the gravity wave
bispectrum was shown in  \cite{Soda:2011am,Shiraishi:2011st}. }. 
Thus for gravitational wave correlators in
$dS$ we only have two possible shapes, both parity conserving.

We show that,
under general principles, the higher derivative corrections can
 be as large as the term that comes from the Einstein
term, though still very small compared to the two point function. In fact, we expect that the ratio
of $\la \gamma  \gamma \gamma \ra /\la \gamma \gamma \ra^2 $ is of order one for the ordinary gravity case, and can
be as big as one for the other shape.  When it becomes one for the other shape it means that the
derivative expansion is breaking down. This happens when the scale controlling the higher derivative
corrections becomes close to the Hubble scale. For example,  the string scale can  get  close
to the Hubble scale. Even though ordinary Einstein gravity is breaking down, we can still compute
this three point function from symmetry considerations, indicating the power of the symmetry based
approach for the three point function.
This gravity three point function
appears to be  outside the reach of the experiments occurring in the near future.
We find it interesting that by measuring the gravitational wave  three point function we can directly
assess the size of the higher derivative corrections in the gravity sector of the theory.
Of course, there are   models of inflation where higher derivatives are important in the
scalar sector \cite{Alishahiha:2004eh, ArmendarizPicon:1999rj}
and in that case too, the non-gaussian corrections are a direct way to test those models
\cite{Garriga:1999vw, Silverstein:2003hf, Chen:2006nt}.
The simplicity of the results we find here is no longer present
when we go from de Sitter to an inflationary
background. However, the results we find are still the leading approximation in the  slow roll
expansion. Once we are
away from the de Sitter approximation, one can still study the higher derivative corrections in a systematic
fashion as explained in \cite{Cheung:2007st, Weinberg:2008hq, Senatore:2010wk}.

Our results have also a ``dual'' use. The computation of the three point function for gravitational waves
is mathematically equivalent to the
computation of the three point function of the stress tensor
in a three dimensional conformal field theory. This is most clear when we consider the wavefunction
of the universe as a function of the metric, expanded around de Sitter space  at late times \cite{Maldacena:2002vr}.
This is a simple consequence of the symmetries, we are not invoking any duality here, but making a
simple statement \footnote{Of course, this statement is consistent with the idea that such a wavefunction can be computed in terms of a dual field theory. Here we are not making any assumption about the
existence of a dual field theory. Discussions of a possible  dual theory in the de Sitter context can be
found in \cite{Strominger:2001pn,Witten:2001kn}. }.
From this point of view it is clear why conformal symmetry restricts the answer. If one were dealing
with scalar operators, there would be only one possible three point function. For the stress tensor, we
have three possibilities, two parity conserving and one parity violating. The parity conserving
three point functions were computed in \cite{Osborn:1993cr}. Here we present these three point functions
in momentum space. Momentum space is convenient to take into account the conservation laws, since
one can easily focus on the transverse components of the stress tensor.
However, the constraints from special conformal
symmetry are a little cumbersome, but manageable. We derived the explicit form of the special
conformal generators in momentum space and we checked that the correlators we computed are the only
solutions.
In fact, we found it convenient to introduce a spinor helicity formalism, which is similar to the one
used in flat four dimensional space.
This formalism simplifies the algebra involving the spin indices  and it is
a convenient way to describe gravitational waves in de Sitter, or stress tensor correlators in a
three dimensional conformal field theory. In Fourier space the stress tensor has a
three momentum $\vec k$, whose square is non-zero. The longitudinal components are determined by
the Ward identities. So the non-trivial information is in the transverse, traceless components.
The transverse
space is two dimensional and we can classify the transverse indices in terms of their helicity.
Thus we have two operators with definite helicity,  $T^\pm(\vec k)$.
In terms of gravitational waves, we are considering gravitational waves that have circular polarization.
These can be described  in a
convenient way by defining two spinors $\lambda$ and $\bar \lambda$, such that $\lambda^{a}
\bar \lambda^{\dot b} = ( \vec k , |\vec k|)^{a \dot b}$. In other words, we form a null four
vector, and we proceed as in the four dimensional case. We only have $SO(3)$ symmetry, rather than
$SO(1,3)$, which allows us to mix dotted and undotted indices. We can then write the polarization
vectors as $\xi^i \propto \sigma^i_{a \dot b}\lambda^a \lambda^{\dot b }$ (no bar), etc.
This leads to simpler expressions
for the three point correlation functions of the stress tensor in momentum space.  We have
expressed the special conformal generator in terms of these variables. One interesting aspect is
that this formalism makes the three point function completely algebraic (up to the delta function for
momentum conservation). As such, it might be a useful starting point for computing higher point functions
in a recursive fashion, both in dS and AdS.
This Fourier representation might also help in the construction of conformal blocks.
The connection between bulk symmetries and the conformal symmetry on the boundary was discussed
 in the inflationary context in
  \cite{Larsen:2002et, Larsen:2003pf, Larsen:2004kf,McFadden:2009fg,McFadden:2010vh}.

The idea of using conformal symmetry to constrain cosmological correlators was also discussed in
 \cite{Antoniadis:2011ib}, which appeared while this paper was in preparation. Though the point of
  view is similar, some of the details differ. In that paper, scalar
 fluctuations were considered. However, scalar fluctuations, and their three point function, crucially
 depend on departures from conformal symmetry. It is likely that a systematic treatment of such a breaking
 could lead also to constraints, specially at leading order in slow roll. On the other hand, the
 gravitational wave case, which is discussed here, directly gives us the leading term in the slow roll expansion.

The paper is organized as follows.
In section \ref{direct} we perform the computation of the most general three point function from a bulk perspective. We also discuss the possible size of  the higher derivative corrections. In section \ref{spinorhelic} we review the spinor helicity formalism in 4D flat spacetime, and propose a similar formalism that is useful for describing correlators of CFTs and expectation values in dS and AdS. We then write the previously computed three point functions using these variables. In section \ref{conformal} we review the idea of viewing
 the wavefunction of the universe in terms of  objects that have  the same symmetries as correlators of
stress tensors in CFT. We also  emphasize how conformal symmetry constrains the possible shapes of the three point function. In section \ref{freeCFT} we explicitly compute the three point function for the stress tensor for free field theories in 3D, and show that, up to contact terms, they have the same shapes as the ones that do not violate parity, computed from the bulk perspective. The appendices contain various technical points and side comments.

{\bf Note added:} We have revised this paper correcting statements regarding 
the parity violating terms in the bispectrum, in light of \cite{Soda:2011am,Shiraishi:2011st}. 

\section{Direct computation of general three point functions}\label{direct}

In this section we compute the three point function for gravitational waves in de Sitter space.
We do the computation in a fairly straightforward fashion. In the next section we will discuss in
more detail the symmetries of the problem and the constraints on the three point function.

\subsection{Setup and review  of the computation of the gravitational wave spectrum }\label{twoptfunction}

The gravitational wave spectrum in single-field slow roll inflation was derived in \cite{Starobinsky:1979ty}.
Here we will compute the non-gaussian corrections to that result. As we discussed above, we will do all
our computations in the de Sitter approximation. Namely, we assume that we have a cosmological
constant term so that the background spacetime is de Sitter.
 There is no inflaton or scalar perturbation in this context. This approximation correctly gives the
leading terms in the slow roll expansion. We leave a more complete analysis to the future.

It is convenient to write the
 metric in the ADM form
\begin{eqnarray}\label{ADMmetric}
g_{00}=-N^2+g_{ij}N^iN^j, & g_{0i}=g_{ij}N^j, & g_{ij}=e^{2 H t}\exp(\gamma)_{ij}
\end{eqnarray}
Where $H$ is Hubble's constant.
 $N$ and $N_i$ are Lagrange multipliers (their equations of motion will not be dynamical), and $\gamma_{ij}$ parametrizes gravitational degrees of freedom.
The action can be expressed as
\begin{equation}\label{ADMaction}
S=\frac{M^2_{Pl}}{2}\int\sqrt{-g} (R-6 H^2)=\frac{M^2_{Pl}}{2}\int \sqrt{g_{3}}\left(N R^{(3)} - 6 N H^2 + N^{-1} (E_{ij}E^{ij} - (E^i{}_i)^2)\right)
\end{equation}

Where $E_{ij}=\frac{1}{2} (\dot g_{ij} - \nabla_i N_j-\nabla_j N_i)$ and we define $M_{Pl}^{-2}\equiv 8\pi G_N$. We fix the gauge by imposing that gravity fluctuations are transverse traceless, $\gamma_{ii}=0$ and $\partial_i\gamma_{ij}=0$. Up to third order in the action,
 we only need to compute the first order values of the Lagrange multipliers $N$ and $N_i$  \cite{Maldacena:2002vr}. By our gauge choice, these are $N=1$ and $N_i=0$ as there
 cannot be a first order dependence on the gravity fluctuations. Expanding the action up to second order in perturbations we find
\begin{equation}\label{2ndorderaction}
S_2=\frac{M^2_{Pl}}{8}\int\left(e^{3 H t} \dot \gamma_{ij} \dot \gamma_{ij}-e^{Ht}\partial_l \gamma_{ij}\partial_l \gamma_{ij}\right)
\end{equation}
We can expand the gravitational waves in terms of polarization tensors and a suitable choice of solutions of the classical equations of motion. If we write
\begin{equation}\label{gammaexpansion}
\gamma_{ij}({\bf x}, t)=\int \frac{d^3k}{(2\pi)^3} \sum_{s=+,-} \epsilon^s_{ij}e^{i{\bf k.x}}\gamma_{cl}(t) a^\dagger_{s,\vec k} +h.c.
\end{equation}
the gauge fixing conditions imply that the polarization tensors are traceless, $\epsilon_{ii}=0$, and transverse $k_i\epsilon_{ij}=0$.  The helicities can be normalized by
$\epsilon^A_{ij}\epsilon^{*B}_{ij}=4\delta^{AB}$.
The equations of motion are then given by
\begin{align}\label{eom2ndorder}
0=&\gamma''_{cl}(\eta)-\frac{2}{\eta}\gamma'_{cl}(\eta)+k^2\gamma_{cl}(\eta)  ~,~~~~~~~~~~~ \eta = - { e^{ - H t } \over H }
\end{align}
where  we introduced conformal time, $\eta$.
We take the classical solutions to be those that correspond to the Bunch-Davies vacuum \cite{Bunch:1978yq}, so $\gamma_{cl}(\eta)=\frac{H}{\sqrt{2k^3}}e^{ik\eta}(1-ik\eta)$.
Here we have denoted by $k = |\vec k|$ the absolute value of the 3-momentum of the wave.
We are interested in the late time contribution to the two-point function, so we take the limit where $\eta\rightarrow0$.
 After Fourier transforming the late-time dependence of the two-point function and
  contracting it with polarization tensors of same helicities we find:
\begin{equation}
\la\gamma^{s_1}_k\gamma^{s_2}_{k'}\ra=(2\pi)^3\delta^3(k+k')\frac{1}{2k^3}\left(\frac{H}{M_{Pl}}\right)^2 4\delta_{s_1s_2}
\end{equation}

In the inflationary context (with a scalar field),
 higher derivative terms could give rise to a parity breaking contribution
to the two point function. This arises from terms in the effective action of the form
$ \int f(\phi) W \widetilde W $ \cite{Lue:1998mq,Alexander:2004wk}. This parity breaking term
leads to a different amplitude for positive and negative helicity gravitational waves, leading to a net
circular polarization for gravitational waves.
If $f$ is constant this term is a total derivative and it   does not
contribute. Thus, in de Sitter   there is no contribution from this term. In other words, the
parity breaking contribution is proportional to the time derivative of $f$.
Some authors have claimed that one can get such parity breaking terms even in pure de Sitter \cite{Contaldi:2008yz}. However, such a contribution would break CPT. Naively, we would expect
that CPT is spontaneously broken because of the expansion of the universe. However, in de Sitter we
can go to the static patch coordinates where the metric is static. For such an observer we expect
CPT to be a symmetry. A different value of the left versus right circular polarization for gravitational
waves would then violate CPT.

In the AdS case, or in a general
CFT,  there can be parity violating contact terms in the two point function\footnote{A contact term is
a contribution proportional to a delta function of the operator positions.}. This is discussed
in more detail in appendix \ref{ParityBreak}.

\subsection{ Three point amplitudes in flat space }

In order to motivate the form of the four dimensional action that we will consider, let us
discuss some aspects of the scattering of three gravitational waves in flat space. This is
relevant for our problem since at short distances the spacetime becomes close to flat space.

In flat space we can consider the on shell scattering amplitude between three gravitational waves.
Due to the momentum conservation condition we cannot form any non-zero Mandelstam invariant from the
three momenta. Thus, all the possible forms for the amplitude are exhausted by listing all the
possible ways of contracting the polarization tensors of the gravitational waves and their momenta, \cite{Benincasa:2007xk}\footnote{In flat space,
the three point amplitude is non-trivial only after analytically
continuing to complex values of the momentum.}.
There are only two possible ways of doing this, in a parity conserving manner.
 One corresponds to the amplitude that comes
from the Einstein action. The other corresponds to the amplitude we would get from a term in the
action that has the form $W^3$, where $W$ is the Weyl tensor.
 In addition, we can write down a parity violating amplitude that
comes  from a term of the form $\widetilde W W^2$, where $\widetilde W_{abcd} =
 \epsilon_{abef} W^{ef}_{~~cd}$.
These   terms involving the Weyl tensor
 are expected to arise from higher derivative corrections in a generic gravity theory.
By using field redefinitions,
any other higher derivative interaction can be written in such a way that it does not contribute to the
three point amplitude.

By analogy, in our de Sitter computation we will consider only the following terms in the gravity
action
\begin{equation}\label{action}
\begin{split}
S_{eff}= \int d^4x{}\left[\sqrt{-g} \left({M_{Pl}^2\over 2}\left(-6 H^2+R \right)+\Lambda^{-2} \left(a~W^{ab}{}_{cd}W^{cd}{}_{mn}W^{mn}{}_{ab}\right)\right)+\right.\\\left.+~\Lambda^{-2} \left(b~\epsilon^{abef}W_{efcd}W^{cd}{}_{mn}W^{mn}{}_{ab}\right)\right]
\end{split}
\end{equation}
Here $\Lambda$ is a scale that sets the value of the higher derivative corrections. We will discuss
its possible values later. This form of the action is enough for generating the most general gravity
three point function that is consistent with de Sitter invariance. This will be shown in more detail
in section \ref{conformal}, by using the action of the special conformal generators.
For the time being we can accept it in analogy to the flat space result.
Instead of the Weyl tensor in \eqref{action} we could have used the Riemann tensor. The disadvantage
would be that the $R^3$ term would not have vanished in a pure de Sitter background and it would also
have contributed to the two point function. However, these extra contributions are trivial and can
be removed by field redefinitions. So it is convenient to consider just the $W^3$ term.

\subsection{Three-point function calculations}\label{3ptcalculation}

In this subsection we compute the three point functions that emerge from the action in (\ref{action}).
First we compute the three point function coming from the Einstein term, and then the one from the $W^3$ term.

\subsection{Three point function from the Einstein term}

This was done in \cite{Arutyunov:1999nw,Maldacena:2002vr}\footnote{In \cite{Arutyunov:1999nw} the $AdS$
case was considered.}.
For completeness,  we review the calculation and give some further details.
To cubic order we can set $N=1$, $N_i =0$ in \eqref{ADMmetric}. Then
 the only cubic contribution from \eqref{ADMaction}
 comes from the term involving the curvature of the three dimensional slices, $ R^{(3)}$.
Let us see more explicitly why this is the case. On the three dimensional slices we define
 $g = e^{ 2 H t} \hat g$, with $\hat g_{ij} = (e^{\gamma})_{ij}$. All indices will be raised and
 lowered with $\hat g$. The action has the form
\begin{equation} \label{ActionR}
S^{(3)}_R={1 \over 2}\int dt  d^3x \left[
e^{ H t } \hat R^{(3)}+ e^{ - H t } (\hat E_{ij}\hat E^{ij} - (\hat E^i{}_i)^2)\right]
\end{equation}
Now we prove that the second term does not contribute any third order term to the action.
To second order in $\gamma$ we have  $\hat E^i_j = ( \dot \gamma + { 1 \over 2 } [\dot \gamma , \gamma ] )_{ij} $. Then we find
\begin{equation} \begin{array}{rcl}
\hat E_{ij}\hat E^{ij} - (\hat E^i{}_i)^2
&=&
 \dot \gamma_{ij} \dot \gamma_{ij} + o(\gamma^4)
\end{array}\end{equation}
which does not have any third order term.
 Thus,  the third order action is  proportional to the  curvature of  the three-metric. This is then
  integrated over time, with the appropriate prefactor in \eqref{ActionR}.
 Of course, if we were doing the computation of the flat space three point amplitude, we could also
 use a similar argument. The only difference would be the absence of the $e^{ H t}$ factor in the
 action \eqref{ActionR}. Thus, the algebra involving the contraction of the polarization tensors and the momenta
 is the same as the one we would do in flat space (in a gauge where the polarization tensors
  are zero in the
 time direction).
  Thus the de Sitter answer is proportional to the
 flat space result, multiplied by a function of $|\vec k_i|$ only, which comes from the fact that the
 time dependent part of the wavefunctions is different.

We want to calculate the tree level three-point function that arises from this third-order action. In order to do that, we use the in-in formalism. The general prescription is that any correlator is given by the time evolution from the ``in" vacuum up to the operator insertion and then time evolved backwards, to the ``in" vacuum again, $\la O(t) \ra=\left\la \text{in}\left| \bar T e^{-i\int H_{int}(t')dt'}O(t) T e^{i\int H_{int}(t')dt'}\right|\text{in}\right\ra$. We are only interested in the late-time limit of the expectation value. We find
\begin{equation}
\la \gamma^{s_1} (x_1,t)\gamma^{s_2} (x_2,t)\gamma^{s_3}(x_3,t)  \ra_{t \rightarrow +\infty} = -i \displaystyle \int_{-\infty}^{+\infty} dt'~[H_{int}(t'),~ \gamma^{s_1} (x_1,+\infty)\gamma^{s_2} (x_2,+\infty)\gamma^{s_3}(x_3,+\infty)]
\end{equation}
We write the gravitational waves in terms of oscillators as in \eqref{gammaexpansion}.
We calculate correlators for gravitons of specific helicities and 3-momenta.
Note that, because there are no time derivatives in the interaction Lagrangian, then it follows that $H^3_{int}=-L^3_{int}$.
Once we put in the wavefunctions, the time integral that we need to compute is of the form $Im[\int_{-\infty}^0 d\eta \frac{1}{\eta^2}(1-i k_1 \eta)(1-ik_2 \eta)(1-i k_3 \eta) e^{i(k_1+k_2+k_3)\eta}]$  (in conformal time).
 Two aspects of the calculation are emphasized here. One is that we need to rotate the contour to damp the exponential factor at early times, which physically corresponds to finding the vacuum of the interacting theory \cite{Maldacena:2002vr},
 as is done in the analogous flat space computation.
    Another aspect is that, around zero, the primitive is of the form $-\frac{e^{i(k_1+k_2+k_3)\epsilon}}{\epsilon}=-\frac{1}{\epsilon} - i(k_1+k_2+k_3)+O(\epsilon)$, where $\epsilon$ is our late-time cutoff. This divergent contribution is real and it drops out from
    the imaginary part. We get
\begin{equation}\label{hhhR}
\begin{split}
\la \gamma^{s_1}_{k_1}  \gamma^{s_2}_{k_2}  \gamma^{s_3}_{k_3} \ra_{R}&= (2\pi)^3 \delta^3 \left({\bf k_1+k_2+k_3}\right) \left(\frac{H}{M_{Pl}}\right)^4\frac{4}{(2k_1k_2k_3)^3} \times \\
& \left[(k^2_ik^2_j \epsilon^1_{ij}) \epsilon^2_{kl}\epsilon^3_{kl} -
 2 \epsilon^1_{ij}(k^3_l \epsilon^2_{li})(k^2_m \epsilon^3_{mj}) + \text{cyclic}\right] \times
 \\
 & \left( k_1+k_2+k_3 - \frac{k_1k_2+k_1k_3+k_2k_3}{k_1+k_2+k_3}-\frac{k_1k_2k_3}{(k_1+k_2+k_3)^2}\right)
\end{split}
\end{equation}
The second line is the one that is the same as in the flat space amplitude. The third line comes
from the details of the time integral.
Below we will see how this form for the expectation value is determined by the de Sitter isometries, or the
conformal symmetry.

\subsection{Three point amplitude from $W^3$ in flat space}
Let us calculate the following term in flat space, to which we will refer as $W^3$: $W^{\alpha \beta}{}_{\gamma \delta} W^{\gamma \delta}{}_{\sigma \rho} W^{\sigma \rho}{}_{\alpha \beta}$.
We can write the following first order expressions for the components of the Weyl tensor
\begin{eqnarray}
W^{0i}{}_{0j}&=&\frac{1}{2} \ddot \gamma_{ij} \nonumber \\
W^{ij}{}_{0k}&=&\frac{1}{2} \left(\dot \gamma_{ki,j}-\dot\gamma_{kj,i}\right)  \nonumber \\
W^{0i}{}_{jk}&=&\frac{1}{2} \left(\dot \gamma_{ik,j}-\dot\gamma_{ij,k}\right) \\
W^{ij}{}_{kl}&=& \frac{1}{2} \left( -\delta_{ik}\ddot \gamma_{jl}+ \delta_{il}\ddot \gamma_{jk}+ \delta_{jk}\ddot \gamma_{il}- \delta_{jl}\ddot \gamma_{ik}\right)  \nonumber
\end{eqnarray}
where we used that $\gamma$ is an on shell gravitational wave.
 i.e. $\gamma$ obeys the flat space equations of motion.  We also used that  $\gamma_{ii}=\partial_i \gamma_{ij}=0$, $N_i=0$, $N=1$.
 We can then write
\begin{equation}
\begin{split}
W^{\alpha \beta}{}_{\gamma \delta} W^{\gamma \delta}{}_{\sigma \rho} W^{\sigma \rho}{}_{\alpha \beta}= W^{ij}{}_{kl} W^{kl}{}_{mn} W^{mn}{}_{ij}+6 W^{0i}{}_{jk} W^{jk}{}_{lm} W^{lm}{}_{0i}+\\+12W^{0i}{}_{0j} W^{0j}{}_{kl} W^{kl}{}_{0i}+8W^{0i}{}_{0j} W^{0j}{}_{0k} W^{0k}{}_{0i}
\end{split}
\end{equation}
Evaluating these terms leads us to
\begin{equation}
S^{(3)}=\displaystyle \int \Lambda^{-2} \left[2 \ddot\gamma_{ij} \ddot\gamma_{jk} \ddot\gamma_{ki} + 3 \ddot \gamma_{ij} \dot \gamma_{kl,i} \dot \gamma_{kl,j} + 3 \ddot \gamma_{ij} \dot \gamma_{ik,l} \dot \gamma_{jl,k} - 6 \ddot \gamma_{ij} \dot \gamma_{ik,l} \dot \gamma_{kl,j} \right]
\end{equation}
Plugging $\gamma_{ij}=\epsilon^1_{ij} e^{\mathrm{i}k_1\cdot x}+\epsilon^2_{ij} e^{\mathrm{i}k_2 \cdot x}+\epsilon^3_{ij} e^{\mathrm{i}k_3 \cdot x}$ where $k\cdot x = k^i x_i - k t$, we get the following expression for the vertex due to the $W^3$ term:
\begin{align}\label{vertflat}
\begin{split}
V_{W^3, flat} &=  6 ~ k_1 k_2 k_3 \left[ k_1~ k_2^i k_2^j \epsilon^1_{ij}  \epsilon^2_{kl}  \epsilon^3_{kl} + \text{cyclic } - \right.
\\&
\left.
(k_1+k_2+k_3)(\epsilon^1_{ij}k_3^k\epsilon^2_{ki} k_2^l \epsilon^3_{lj}+ \text{cyclic}) -2 ~ k_1 k_2 k_3\epsilon^1_{ij} \epsilon^2_{jk} \epsilon^3_{ki}  \right]
\end{split}
\end{align}
By choosing a suitable basis for the polarization tensors, one can show that this agrees with the gauge invariant covariant expression
$V_{W^3, flat}= 6 k_1^\mu k_1 ^\nu \epsilon^1 _{\rho \sigma} k_2 ^\rho k_2^\sigma \epsilon^2_{\eta \tau} k_3^\eta k_3^\tau \epsilon^3_{\mu\nu}$.

\subsection{Three point function from $W^3$ in dS}

The straightforward way of performing the computation would be to insert now the expressions for
the wavefunctions in the $W^3$ term in de Sitter space, etc.
There is a simple observation that allows us to perform the de Sitter computation. First we observe
that the Weyl tensor is designed so that it transforms in a simple way under overall Weyl rescaling
of the metric. Thus the Weyl tensor for the metric in conformal time is simply given by
$W_{\mu \nu \delta\sigma}(g) = { 1 \over H^2 \eta^2 } W_{\mu \nu \delta \sigma}(\hat g = e^\gamma)$.
Note also that, for this reason, the Weyl tensor vanishes in the pure de Sitter background. Thus,
 we only need to evaluate the Weyl tensor at linearized order\footnote{Note that the $W^3$ term does not
 contribute to the two point function.}.
For on shell wavefunctions $\gamma =  ( 1 - i k \eta) e^{ i k \eta + i \vec k . \vec x }$ we can show that
\begin{equation}
W_{\mu \nu \delta \sigma}( \gamma) =  -i  |\vec k| \eta W^{flat}_{\mu \nu \delta \sigma}(e^{i k \eta + i \vec k . \vec x } )
\end{equation}
where $W^{flat}$ is the expression for the flat space Weyl tensor that we computed in the previous
section, computed at linearized order for a plane wave around flat space.
 Thus, when we insert these expressions in the action we have
 \begin{equation}
 S = \int  W^3 = \int_{-\infty}^0  d\eta d^3 x   ( k_1k_2k_3 \eta^3)  (H^2 \eta^2 ) ( W^{flat} )^3
 \end{equation}

The whole algebra involving polarization tensors and momenta is exactly the same as in flat space.
The only difference is the time integral, which now involves a factor of the form $\int d\eta \eta^5 e^{ i E \eta } \propto 1/E^6$, where we have defined $E = k_1 + k_2 + k_3$, and we rotated the contour
appropriately.
Putting all this together, we get the following result for the three-point function due to the $W^3$ term
\begin{equation}\label{hhhW3dS}
\begin{split}
\la \gamma^{s_1}_{k_1}  \gamma^{s_2}_{k_2}  \gamma^{s_3}_{k_3} \ra_{W^3} &
=(2\pi)^3\delta^3 \left({\bf k_1+k_2+k_3}\right) \times
\\\
& \left(\frac{H}{M_{Pl}}\right)^6\left(\frac{H}{\Lambda}\right)^2 a \frac{(-30)}{(k_1+k_2+k_3)^6(k_1k_2k_3)^2}V_{W^3, flat}
\end{split}
\end{equation}
where $V_{W^3,flat}$ was introduced in \eqref{vertflat}. There are also
 factors of $1/k^3_i$  that  were included to get this result.
The parity violating piece will be discussed after
 we introduce spinor variables, because they will make the calculation much simpler.

\subsection{Estimating the size of the corrections}

Let us write the effective action in the schematic form
\begin{equation}\label{acts}{
S = {M^2_{Pl} \over 2 } \left[  \int [\sqrt{g} R -6  H^2 \sqrt{g} ]   + L^4  \int W^3 \right] + \cdots
}
\end{equation}
where the dots denote other terms that do not contribute to the three point function. Here $L$ is a
constant of dimensions of length.
We have pulled out an overall power of $M^2_{Pl}$ for convenience. 
The gravitational wave expectation values coming from this Lagrangian have the following orders of magnitude
\begin{equation}
\la \gamma \gamma \ra \sim { H^2 \over M_{Pl}^2 } ~,~~~~~~~ \la \gamma \gamma \gamma \ra_R = {H^4 \over M_{Pl}^4 }  ~~~~~~~~ \la\gamma \gamma \gamma \ra_{W^3} = {H^4 \over M_{Pl}^4} ( L H )^4
\end{equation}
 Thus the ratio between the two types of non-gaussian  corrections is
 \begin{equation}\label{correcta}{
  { \la \gamma \gamma \gamma \ra_{W^3} \over
\la \gamma \gamma \gamma \ra_R } \sim  {  L^4 H^4 }
  }
  \end{equation}
We know that $H^2/M^2_{Pl}$ is small. This parameter controls the size of the fluctuations.
 In the $AdS$ context, we know that when the right hand side in (\ref{correcta})
 becomes of order one we have
 causality problems \cite{Hofman:2008ar, Brigante:2007nu, Brigante:2008gz, Hofman:2009ug}.
We expect that  the same is true in $dS$, but we have not computed the precise value
of the numerical coefficient where such causality violation would occur.  So  we expect that
 \begin{equation}\label{unitbound}{
 { H L  } \lesssim 1
 }
 \end{equation}
In a string theory context we expect $L$ to be of the order of the string scale, or the Kaluza Klein
scale. Thus the four dimensional gravity description is appropriate when $H L \ll1$. In fact, in
string theory we expect important corrections when $H \ell_s \sim 1$. In that case, the string
length is comparable to the Hubble scale and we expect to have important stringy corrections
to the gravity expansion. Note that in the string theory context we can still have $H^2/M_{Pl}^2 \sim g_s^2 $ being quite small. So we see that there are scenarios where the higher derivative corrections are as
important as the Einstein contribution, while we still have a small two point function, or small
expansion parameter $H^2/M_{Pl}^2$. In general, in such a situation we would not have any good
argument for neglecting higher curvature corrections, beyond the $W^3$ term. However, in the
particular case of the three point function, we can just consider these two terms and that is enough, since
these two terms (the Einstein term and the $W^3$ term) are enough to parametrize all the possible three
point functions consistent with de Sitter invariance.
If we define an $f_{NL-gravity} = \la\gamma \gamma \gamma \ra/\la \gamma \gamma \ra^2$, then we find that the
Einstein gravity contribution of $f_{NL-gravity}$ is of order one. This is in contrast to the $f_{NL}$ for
scalar fluctuations which, for the simplest models, is suppressed by an extra slow roll factor\footnote{
Note that we have divided by the {\it gravity} two point function to define $f_{NL-gravity}$. If we had
divided by scalar correlators, we would have obtained a factor of $\epsilon^2$. }.

In an inflationary situation we know that the fact that the fluctuations are small is an indication that
the theory was weakly coupled when the fluctuations were generated. However, it could also be that
the stringy corrections, or higher derivative corrections were sizable. In that case, we see that the
gravitational wave three point function (or bispectrum)
gives a direct measure of the size of higher derivative corrections. Other ways of trying to see these
corrections, discussed in \cite{Kaloper:2002uj}, involves a full reconstruction of the potential, etc. In an inflationary
context terms involving the scalar field and its time variation could give rise to new shapes for
the three point function since conformal symmetry would then be broken. However, one expects
such terms to be suppressed by slow roll factors relative to the ones we have considered here.
However a model specific analysis  is  necessary to see whether   terms that
contain slow roll factors, but less powers of $L H$   dominate over the ones we discussed.
For example, a term of the form $M^2_{Pl} L^2 f(\phi) W^2$ is generically present in the
effective action\cite{Weinberg:2008hq}. Such a term could give a correction of the order $\la \gamma \gamma \gamma\ra_{f W^2} /\la \gamma \gamma \gamma \ra_R  \sim  \epsilon_f (H L)^2$. Here $\epsilon_f$ is a small quantity of the order of a
 slow roll parameter, involving the time derivatives of $f$.
 Whether this dominates or not relative to
(\ref{correcta}) depends on the details of the inflationary scenario. In most cases, one indeed
expects it to dominate. It would be very interesting if (\ref{correcta}) dominates because it is a direct
signature of higher derivative corrections in the {\it gravitational} sector during inflation.

Notice that the upper bound (\ref{unitbound}) is actually smaller than the naive expectation
from the point of view of the validity of the effective theory. From that point of view we would
simply demand that the correction due to $W^3$ at the de Sitter scale $H$ should be smaller than one.
This requires the weaker bound $ H^4 L^4 < { M_{Pl} \over H } $.
This condition is certainly too lax in the $AdS$ context, where one can argue for the more restrictive
condition \eqref{unitbound}.

   In summary,
  we can make the higher derivative  contribution to the gravity three point function
 of the same order
  as the Einstein Gravity contribution. Any of these two terms are, of course, fairly small to
begin with.

\section{Spinor helicity variables for de Sitter computations}\label{spinorhelic}

In this section we introduce a technical tool that simplifies the description of gravitons
in de Sitter. The same technique works for anti-de Sitter and it can also be applied for conformal
field theories, as we will explain later.

The spinor helicity formalism is a convenient way to describe scattering amplitudes of
massless particles with spin in
four dimensions.  We review the basic ideas here. For a more detailed description, see \cite{Witten:2003nn,Benincasa:2007xk,ArkaniHamed:2008gz,Cheung:2009dc}.
In four dimensions the Lorentz group is $SO(1,3)\sim SL(2) \times SL(2)$.
A vector such as $k_\mu$ can be viewed as having two $SL(2)$ indices, $k^{a \dot b }$.
The new indices run over two values.
  A 4-momentum that obeys the mass shell condition, $k^2 =0$ can be represented as
 a product of two (bosonic)
 spinors $k^{a \dot b } = \lambda^a \bar \lambda^{\dot b}$.  Note that if we rescale $\lambda \to w \lambda$ and $\bar \lambda \to { 1 \over w }  \bar \lambda$ we get the same four vector. We shall
 call this the ``helicity'' transformation.
  Similarly, the polarization vector
 of a spin one particle $\xi_\mu$ with negative helicity can be represented as
 \begin{equation}  \label{polari}
 \xi^-{}^{a \dot b } = \frac{ \lambda^a \bar \mu^{\dot b } }{ \la \bar \lambda , \bar \mu \ra }
\end{equation}
where we used the $SL(2)$ invariant contraction of indices $\la \lambda , \mu\ra \equiv \epsilon_{ab}\lambda^a \mu^b $, where $\epsilon_{ab}$ is the SL(2) invariant epsilon tensor.
We have a similar tensor $\epsilon_{\dot a \dot b}$ to contract the dotted indices. We cannot
contract an undotted index with a dotted index.
 Note that this polarization vector \eqref{polari} is not invariant under
the helicity transformation. In fact, we can assign it a definite helicity weight, which we call minus one. This  polarization tensor \eqref{polari} is independent of
the choice of $\bar \mu$. More precisely,  different choices of $\bar \mu$
correspond to gauge transformations on
the external particles. For negative helicity we exchange $\lambda , \bar \eta \leftrightarrow \bar \lambda, \eta $ in \eqref{polari}.  For the graviton we can write the polarization tensor as a ``square''
of that of the vector
\begin{eqnarray}\label{polariznvecs}
\xi^{+}{}^{ab\dot a \dot b}=\frac{\mu^a \mu^b \bar \lambda^{\dot a} \bar \lambda^{\dot b}}{\la \mu,  \lambda \ra^2}, && \xi^{-}{}^{ab\dot a \dot b}=\frac{\lambda^a \lambda^b \bar \mu^{\dot a} \bar \mu^{\dot b}}{\la \bar \lambda, \bar \mu  \ra^2}
\end{eqnarray}
The product of two four vectors can be written as $ k . k' = - 2 \la \lambda ,\lambda'\ra \la \bar \lambda ,
\bar \lambda' \ra$.

Now let us turn to our problem. We are interested in computing properties of gravitational waves at late time.
We still have the three momentum $\vec k$. This is not null. However, we can just define a null
four momentum $( |\vec k| , \vec k)$. This is just a definition. We can now introduce $\lambda $ and
$\bar \lambda$ as we have done above for the flat space case.
In other words, given a three momentum $\vec k$ we define $\lambda $, $\bar \lambda $ via
\be
\label{defllb}
( |\vec k| , \vec k )^{a \dot b} = ( |k| \, \sigma^{0\,a \dot b} + \vec k . \vec \sigma^{a \dot b } ) =  \lambda^a \bar \lambda^{\dot b}
\ee
In the de Sitter problem we do not have full $SL(2) \times SL(2)$ symmetry. We only have one $SL(2)$
symmetry which corresponds to the $SO(3)$ rotation group in three dimensions. This group is diagonally
embedded into the $SL(2)\times SL(2)$ group we discussed above. In other words, as we perform a spatial
rotation we change both the $a$ and $\dot a $ indices in the same way. This means that we now
have one more invariant tensor, $\epsilon_{\dot b a}$ which allows us to contract the dotted with the
undotted indices. For example, out of $\lambda^a$ and $\bar \lambda^{\dot b}$ we can construct
$\la \lambda , \bar \lambda \ra $ by contracting with $\epsilon_{ \dot b a}$. This is proportional to
$|\vec k|$. Thus, this contraction is equivalent to picking out the zero component of the null vector.
When we construct the polarization tensors of gravitational waves, or of vectors, it is convenient to choose them so that their zero component vanishes. But, we have already seen that extracting the zero component involves contracting dotted and undotted indices.
We can now then choose a special $\bar \mu$ in \eqref{polari} which makes sure that the zero component
vanishes. Namely, we choose $\bar \mu^{\dot b} = \lambda^b$. This would not be allowed under the four
dimensional rules, but it is perfectly fine in our context.
 In other words, we choose polarization
 vectors  of the form
\begin{eqnarray} \label{polVec}
 \xi^+{}^{a \dot b} =  \frac{ \bar \lambda^a \bar \lambda^{\dot b } }{ \la \bar \lambda , \lambda \ra}~,~~~~~~~~~~\xi^-{}^{a \dot b } = \frac{ \lambda^a \lambda^{\dot b } }{ \la  \lambda , \bar \lambda \ra }
\end{eqnarray}
Notice that the denominator is just what we were calling $k = |\vec k|$. Note also that the zero component of $\xi$ is zero, since this involves contracting the $a$ and
$\dot b$ indices. This gives a vanishing result due to the antisymmetry of the inner product.
In our case we have a delta function for momentum conservation due to translation invariance, but we do not have one for energy
conservation. The delta function for momentum conservation can be written by contracting
$\sum_I \lambda_I^a \bar \lambda_I^{\dot b}$ with $\sigma^{i \, a \dot b} $ in order to get the spatial
momentum. Alternatively we can say that $ \sum_I \lambda_I^a \bar \lambda_I^{\dot b} \propto \epsilon^{a \dot b } $. This is just saying that the fourvector has only a time component.

For the graviton, we likewise take $\mu = \bar \lambda$ and $\bar \mu = \lambda$ in \eqref{polariznvecs}. With these choices we make sure that the polarization vector has zero time
components and that it is transverse to the momentum.

Everything we said here also applies for correlation function of the stress tensor in three dimensional
field theories. If we have the stress tensor operator $T_{ij}(k)$ in Fourier space, we can then
contract it with a polarization vector transverse to $k$ constructed from $\lambda$ and $\bar \lambda$.
In other words, we construct operators of the form $T^{+} = \xi^+_i \xi^+_j T_{ij}$ with
$\xi^+$ as in \eqref{polVec}.
This formalism applies for any case where we have a four dimensional bulk and a three dimensional
boundary, de Sitter, Anti-de Sitter, Hyperbolic space, Euclidean boundary, Lorentzian boundary, etc.
The only difference between various cases are the reality conditions.
For example, in the de Sitter case that we are discussing now, the reality condition is
$ (\bar \lambda^{\dot a})^* = \epsilon_{\dot a b} \lambda^b $.

In summary, we can use the spinor helicity formalism tyo describe gravitational waves in de Sitter, or
any inflationary background. It is a convenient way to take into account the rotational symmetry
of the problem. One can rewrite the expressions we had above in terms of these variables.

\subsection{Gravitational wave correlators in the spinor helicity variables}

Let us first note the form of the two point function. The only non-vanishing two point functions are
the $++$ and $--$ two point functions. This is dictated simply by angular momentum conservation along
the direction of the momentum. Since the momenta of the two insertions are opposite to each other, their
spins are also opposite and sum to zero as they should.
The two point functions are then
\begin{equation}
 \la \gamma^+ \gamma ^+ \ra =
 \delta^3( k + k') { \la \lambda , \lambda' \ra^2 \over \la \lambda , \bar \lambda \ra^5 } = \delta^3( k + k') {1 \over \la \lambda , \bar \lambda \ra^3  }
 \end{equation}
 where in the last formula we have used a particular expression for $\lambda'$ in terms of $\bar \lambda$. More precisely, if the momentum of one wave if $\vec k$, with its associated $\lambda$ and $\bar \lambda$, then
 for  $\vec k' = - \vec k$ we can choose  $\lambda' = \bar \lambda$ and
$\bar \lambda' = - \lambda$.
Here we have used that the matrices $\sigma^{i}_{ \, a \dot b}$ are symmetric. In the first expression we can clearly see the helicity weights of the expression.
For the $--$ one we get a similar expression.

We can now consider the three point functions. The simplest to describe are the ones coming from the
 $W^3$ interaction. In fact, these contribute only to the $+++$ and $---$ correlators, but not
 to the $++-$ correlators. This is a feature which is also present in the flat space case.
 These non vanishing correlators can be rewritten as 
\begin{align}\label{Wsp}
\la \gamma^{+}_{k_1}  \gamma^{+}_{k_2}  \gamma^{+}_{k_3} \ra_{ W^3}&= {\cal M}  \frac{(-2^8\times3^2\times5)}{(k_1+k_2+k_3)^6(k_1k_2k_3)^2}   [\la \bar 1,\bar 2 \ra\la \bar 2, \bar 3 \ra\la \bar 3, \bar 1 \ra]^2\nonumber \\
\la \gamma^{-}_{k_1}  \gamma^{-}_{k_2}  \gamma^{-}_{k_3} \ra_{ W^3}&={\cal M}  \frac{(-2^8\times3^2\times5)}{(k_1+k_2+k_3)^6(k_1k_2k_3)^2}[\la 1,2 \ra\la 2, 3 \ra\la 3,1 \ra]^2
\\
& {\cal M}  =\displaystyle (2\pi)^3\delta^3 \left({\bf k_1+k_2+k_3}\right) \left(\frac{H}{M_{Pl}}\right)^6\left(\frac{H}{\Lambda}\right)^2\nonumber
\end{align}
where the $k_n$ in the denominators can also be written in terms of brackets such as $k_n = - \langle n , \bar n \rangle$, if so desired. Note that, when rewritten in terms of the $\lambda_n$ and $\bar \lambda_n$,
the above expressions are just rational functions of the spinor helicity variables (up to the overall momentum conservation delta function).
One can check   that indeed the $+ + -$ and $- - +$ vertices vanish for the $W^3$ term, which is straightforward by using the expressions in appendix \ref{spinexp}. Note that this is not trivial because we do not have four-momentum conservation, only the three-momenta are conserved.
The parity violating interaction $ W^2 \widetilde W $ does not contribute to the de Sitter
expectation values \cite{Soda:2011am,Shiraishi:2011st}. 

The Einstein term contributes to all polarization components
\begin{align} \label{ppp}
\la \gamma^{+}_{k_1} \gamma^{+}_{k_2} \gamma^{+}_{k_3} \ra_R &= (2 \pi)^3 \delta^3 \left(\sum_i k_i\right) \left(\frac{H}{M_{Pl}}\right)^4\frac{2}{(k_1k_2k_3)^5}\left[ (k_1+k_2+k_3)^3 -\right. \nonumber \\ &\left.-(k_1k_2+k_1k_3+k_2k_3)(k_1+k_2+k_3)-k_1k_2k_3\right] \left[ \la \bar 1, \bar 2 \ra \la \bar 2, \bar 3 \ra \la \bar 3, \bar 1 \ra\right]^2\\ \label{PPm}
\la \gamma^{+}_{k_1} \gamma^{+}_{k_2} \gamma^{-}_{k_3} \ra_R &= (2 \pi)^3 \delta^3 \left(\sum_i k_i \right) \left(\frac{H}{M_{Pl}}\right)^4\frac{1}{8(k_1k_2k_3)^5} \left[(k_1+k_2-k_3)(k_1-k_2+k_3)(k_2+k_3-k_1)\right]^2 \nonumber \\& \left( k_1+k_2+k_3 - \frac{k_1k_2+k_1k_3+k_2k_3}{k_1+k_2+k_3}-\frac{k_1k_2k_3}{(k_1+k_2+k_3)^2}\right) \left[\frac{\la \bar 1,\bar 2 \ra^3}{\la \bar 1, \bar 3 \ra\la \bar 3, \bar 2 \ra} \right]^2
\end{align}
and similar expressions for $--+$ and $---$.
Note that the Einstein gravity contribution to $+++$ or $---$ is non-vanishing.
This is in contradistinction to what happens in flat space, where it does not contribute to the $+++$ or
$---$ cases. This might seem surprising, given that we had said before that the polarization tensor
contribution to the time integrand is  the same as the flat space one.
After doing the time integral, in flat space we get
  energy conservation, which we do not have here. This explains why we got a non-vanishing answer.
In fact,
 the flat space amplitude is recovered from the above expressions by focusing on the coefficients
of the double poles in $E = k_1 + k_2 + k_3$. The fact that \eqref{ppp} does not have a double pole ensures
that the flat space answer is zero for those polarizations\footnote{ In comparing to the flat space
result, there are also factors of $(k_1k_2k_3) $ that come
from the normalization of the wavefunction. }. Similarly, the flat space answers for $W^3$ are obtained
by looking at the coefficient of the  $6^{th}$ order pole in $E$ in \eqref{Wsp}.

The expressions \eqref{ppp} can also be written in a form that shows explicitly the effect of changing the helicity of one particle:
\begin{align}\label{newform}
&\la \gamma^{+}_{k_1} \gamma^{+}_{k_2} \gamma^{+}_{k_3} \ra_R = \mathcal{N} (k_1+k_2+ k_3)^2  (\la \bar 1, \bar 2 \ra \la \bar 2, \bar 3 \ra \la \bar 3, \bar 1\ra)^2\\
&\la \gamma^{+}_{k_1} \gamma^{+}_{k_2} \gamma^{-}_{k_3} \ra_R = \mathcal{N} (k_1+k_2- k_3)^2  (\la \bar 1, \bar 2 \ra \la \bar 2, 3 \ra \la 3, \bar 1 \ra)^2\\
\mathcal{N}&=(2 \pi)^3 \delta^3 \left(\sum_i k_i\right) \left(\frac{H}{M_{Pl}}\right)^4\frac{2}{(k_1k_2k_3)^5} \times \nonumber\\ &\times \left( k_1+k_2+k_3 - \frac{k_1k_2+k_1k_3+k_2k_3}{k_1+k_2+k_3}-\frac{k_1k_2k_3}{(k_1+k_2+k_3)^2}\right)
\end{align}

In the next section we will show that
 the forms of these results follow from demanding conformal symmetry.

\section{Gravitational wave correlation function and conformal symmetry}\label{conformal}

In this section we will show how the three point functions we discussed above are constrained
by conformal symmetry.

\subsection{Wavefunction of the universe point of view}

In order to express the constraints of conformal symmetry it is convenient to take
the following
  point of view on the computation of the gravity expectation values.
Instead of computing expectation values for the gravitational waves, we can compute the probability
to observe a certain gravitational wave, or almost equivalently, the wavefunction $\Psi( \gamma)$. The expectation
values are given by simply taking $|\Psi(\gamma)|^2 $  and integrating over $\gamma$.
This point of view is totally equivalent to the usual one, where one computes expectation values
of $\gamma$. It is useful because it makes the connection to $AdS$ very transparent\footnote{In fact, the perturbative de Sitter computation is simply an analytic continuation
of the perturbative Anti-de Sitter computation \cite{Maldacena:2002vr}.}. It also makes
the action of the symmetries more similar to the action of the symmetries in a conformal field theory.
This is explained in more detail in \cite{Maldacena:2002vr} (see also \cite{deBoer:1999xf}).

One writes the wavefunction in the form:
\begin{equation}
\begin{split} \label{WFU}
\Psi&=\exp \left(\frac{1}{2}\int d^3x d^3y \la T^s(x)T^{s'}(y) \ra \gamma^s(x)\gamma^{s'}(y) +\right.\\
&\left.+\frac{1}{6}\int d^3x d^3y d^3z\la T^s(x)T^{s'}(y)T^{s''}(z) \ra \gamma^s(x)\gamma^{s'}(y)\gamma^{s''}(z)  + ~\cdots \right)
\end{split}
\end{equation}
The first term expresses the simple fact that the wavefunction is gaussian. From this point of view,
the
quantity  $\la T^s(x)T^{s'}(y) \ra$ is just setting the variance of the gaussian. Namely, this is just
a convenient name that we give to this variance. Similarly for the cubic term, which
 is responsible for the first non-gaussian correction, etc. Here we have ignored
  local terms that are purely imaginary and which drop out when we take the absolute value
 of the wavefunction.
 From this expression for
the wavefunction one can
  derive the following forms for the two and three point functions \cite{Maldacena:2002vr}, to
leading order in the loop expansion, 
\begin{align}
\la\gamma^{s_1}_{k_1}\gamma^{s_2}_{k_2}\ra&=-\frac{1}{2 \la T^{s_1}_{k_1}T^{s_2}_{k_2}\ra}\\ \label{threeTran}
\la\gamma^{s_1}_{k_1}\gamma^{s_2}_{k_2}\gamma^{s_3}_{k_3}\ra&=-\frac{  \la T^{s_1}_{k_1}T^{s_2}_{k_2}T^{s_3}_{k_3}\ra  +\la T^{s_1}_{-k_1}T^{s_2}_{-k_2}T^{s_3}_{-k_3}\ra ^*  }{ \Pi_i (2 \la T^{s_i}_{k_i}T^{s_i}_{-k_i}\ra)}
\end{align}

So we see that it is easy to go from  the description in terms of a wavefunction to
the description in terms of expectation values of the metric. The complex conjugate arises
from doing $|\Psi|^2$ and  we used that $\gamma^s(-\vec k)^* = \gamma^{s}(\vec k)$.
However, if the wavefunction contains terms that are pure phases, we can loose this information
when we consider expectation values of the metric. Precisely this happens when we have the
parity violating interaction $\int W^2 \widetilde W$. It contributes to a term that is a pure phase.

Here $\Psi$ is the usual Wheeler de Witt wavefunction of the universe, evaluated in perturbation theory.
It is expressed in a particular gauge, because we have imposed the $N=1$, $N_i=0$ conditions.
The usual reparametrization constraints and Hamiltonian constraints boil down to some identities on
the functions appearing in \eqref{WFU}. These identities are precisely the Ward identities obeyed
by the stress tensor in a three dimensional conformal field theory\footnote{Though the Ward identities are
the same, some of the positivity constraints of ordinary CFT's are not obeyed. For example, the
$\langle T T \rangle$ two point function is negative. Thus, if there is a dual CFT, it should have
this unusual property. }. In the $AdS$ case, this is
of course familiar from the $AdS/CFT$ point of view. In the de Sitter case, it is also true since
this wavefunction is a simple analytic continuation of the $AdS$ one. It is an analytic continuation
where the radius is changed by $i$ times the radius. In any case, one can just derive directly
these Ward identities from the constraints of General Relativity.
These  identities express the fact that the wavefunction is reparametrization invariant.
For the case that we have scalar operators (and corresponding scalar fields in $dS$) we
get an identity of the form $ \partial_i \la T_{ij}(x) \prod_{k}O(x_k) \ra =
-\sum_l \delta^3(x -x_l) \partial_{x^j_l} \la \prod_k O(x_k)\ra$.  These are derived by starting
with the reparametrization constraint, taking multiple derivatives with respect to the arguments of
the wavefunction, and setting all fluctuations to zero after taking the derivatives.
  There is also another identity coming from the Hamiltonian constraint. This involves
 the trace of $T$ and it takes  into account the   dimension
  of the operator. Namely we have  $ \la T_{ii}(x) \prod_{k}O(x_k) \ra =
-\sum_l \delta^3(x -x_l) \Delta_l  \la \prod_k O(x_k)\ra$.  From these two identities, we can derive equations for the correlation functions
  if we have a conformal Killing vector \footnote{
   If $v^j$ is the conformal Killing vector, then we can multiply
  the $\partial_i T_{ij} \cdots $ equation
   by $v^j$, integrate over $x$, integrate by parts, use the conformal Killing vector equation
$ \partial_{(i} v_{j)} = { 1 \over 3 } \eta_{ij} ( \partial . v )$, use the  $T_{ii}$ equation
 and obtain the equations
$  \sum_s [  v^i(x_s)  \partial_{x^i_s} +  { \partial . v (x_s) \over 3 } \Delta_s ]  \langle \prod_{k}O(x_k) \ra =0$. These encode all the equations obeyed by correlators that are a consequence
of the de Sitter isometries at late times. For the metric, or stress tensors, the equations
contain more indices and we write them in detail below.}. Alternatively, we can derive these
equations simply by noticing that a conformal reparametrization does not change the metric on the
boundary, up to a rescaling (or a shift of time in the bulk). Thus, this leaves the wavefunction
explicitly invariant, without even changing the metric, which is why we get equations on correlation
functions for each isometry of the background space. From the general relativity point of view, this
is just the statement that each isometry of the background leads to a constraint on the wavefunction.
In our  case, the operators  are other insertions
of the stress tensor. Thus, we can think of the coefficients $\langle TT \rangle $ and $\langle TTT\rangle$
appearing in \eqref{WFU} as correlation functions of ``stress tensors''. We are not assuming the
existence of a dual CFT, we are simply saying that these quantities obey the same Ward identities
as the ones for the stress tensor in a CFT. A more precise discussion of these identities can be
found in appendix \ref{wardcheck} and in section \ref{SpecConf}.

The isometries of de Sitter translate into symmetries of the wavefunction. Some of these are simple,
like translation invariance. A less trivial one is dilatation invariance, or scale invariance.
This simply determines the overall scaling of the three point function in terms of the momentum.
If we think of $\gamma$ as a dimensionless variable, then its Fourier components have dimension
minus three. Thus the total dimension of any $n$ point function is $- 3 n$. The delta function of
momentum conservation takes into account a $-3$, and the remainder is the overall degree of homogeneity in
the momentum. For the two point function it is $-3$ and for the three point function it is $-6$. It is
a simple matter to count powers of momenta in the expressions we have given in order to check that
this is indeed the case.

If instead we look at correlators of the stress tensor, then in position space, we have that
the operator has dimension three, while in momentum space it has dimension zero.

The constraints from special conformal transformation are harder to implement and we discuss them
in the next section. The results of the following section are also valid in any three dimensional
conformal field theory. The general form for the three point function in position space
was given in \cite{Osborn:1993cr}. Here we study the same problem in momentum space. The expressions
we find seem a bit simpler to us than the ones in \cite{Osborn:1993cr}, but the reader can judge by
him or herself.

\subsection{Constraints from special conformal transformations}
\label{SpecConf}

de Sitter space is invariant under a full $SO(1,4)$ symmetry group. The metric $ds^2  = { - d \eta^2 + dx^2 \over \eta^2 }$ makes some of these isometries manifest. In particular the scaling symmetry changes
$x_i \to \alpha  x_i$ and $\eta \to \alpha \eta $. There are also three more isometries that are given
in infinitesimal form by
\be
 x^i \to x^i + b^i ( - \eta^2 + \vec x . \vec x ) -2 x^i (\vec b . \vec x ) ~,~~~~~ \eta \to \eta - 2 \eta (\vec b . \vec x )
 \ee
where $\vec b$ is infinitesimal. When $\eta \to 0$, which is the future boundary of the space, the time
rescaling acts in a simple way on the wavefunction. In addition we can drop the $\eta^2$ in the space part. The transformation then becomes what is called a ``special conformal'' transformation on the boundary
parametrized by $\vec x$. In this section we study the action of these transformations in detail.

Now we work with coordinates on the three dimensional boundary.
A special conformal transformation is given by
\begin{align}\label{changeva}
\delta x^i  &=  x^2 b^i - 2 x^i ( x.b)
\\
\Sigma^i_{~~j} &\equiv { \partial \delta x^i \over \partial  x^j } =
 2 ( x^jb^i  - x^i b^j ) - 2 \delta^i_j ( x . b) \equiv
2 \widehat M^j{}_i  - 2 \delta^i_j  ( x . b ) \label{hatMdef}
  \\
  J & =  \det ( 1 + \Sigma ) ^{1/3}  \sim  1 + { 1 \over 3} \sum_{\nu =1}^3  \Sigma^\nu_{~~\nu} = 1 - 2 ( x. b )
\end{align}
where $b^i$ is an infinitesimal parameter.
  The transformation law for a tensor is
  \begin{equation}\label{transfte}
  T'_{\nu'_1 \cdots \nu'_{n} } (x') = { 1 \over J^{\Delta - n}} \left({ \partial x^{\sigma_1} \over \partial {x'}^{\nu'_1} } \cdots { \partial x^{\sigma_n} \over \partial {x'}^{\nu'_n} } \right)T_{\sigma_1 \cdots \sigma_n} (x)
  \end{equation}
  where $\Delta$ is the conformal dimension.
  For a current or the
  stress tensor we have   $\Delta =2,3$ respectively. These transformation laws describe the (infinitesimal) action of
  the de Sitter isometries on the comoving coordinates at late time.

  Now, in order to compute the variation of a correlator, we are interested in its change as a function.
  This means that the transformed correlator, as a function of the new variables, $x'$ should be the
  same as the old correlator as a function of $x$.
  Thus we can evaluate $T'(x)$ (and not $x'$). Then we write $x = x' - \delta x$.
  In that way we find that the change is
 \begin{align}\label{changeco}
  \delta T_{\sigma_1 \cdots \sigma_n} =&  \Delta  2 ( x.b)T_{\sigma_1 \cdots \sigma_n}  - 2  \sum_{ l=1}^n  \widehat M^{\nu_l}_{~~\sigma_l}
   T_{\sigma_1 \cdots \nu_l \cdots \sigma_n} - D T_{\sigma_1 \cdots \sigma_n}
   \cr
   & D \equiv x^2  ( b . \partial) - 2 ( b . x ) ( x . \partial )
   \end{align}
  The matrix $\widehat M$  was defined in \eqref{hatMdef}.
 We now   Fourier transform  \eqref{changeco} .
  The terms that contain a single power of $x$ are easy to transform. They are given simply by
  inserting factors of $x \to -i \partial_k$. For the term involving a $D$, it is important that we first
  replace the $x \to -i \partial_k$ and {\it then} we change the derivatives by factors of $k$,
  $\partial_x
  \to - i k $.
  Thus a term like
  \begin{align}\label{sch}
   x^2 \partial_i \to&  i  {\vec \partial}_k^2  k_i = i ( k_i { \vec \partial}_k^2 + 2 \partial_{k^i} )
   \\
   x^i ( x . \partial_x ) \to &   i ( \partial_{k^i} ( \partial_{k^j} k_j ) ) =
   i [ 4 \partial_{k^i}  + k_j \partial_{k_j} \partial_{k^i } ]
   \end{align}

   We then find that the special conformal generator (up to an overall $i$), now has the form
\begin{align}\label{changep}
  \delta T_{i_1 \cdots i_n}(k) = &
     - (\Delta -3)  2 ( b .\partial_k)T_{i_1 \cdots i_n}(k) + 2 \sum_{ l=1}^n  \widetilde  M^{j_l}_{~~i_l}
   T_{i_1 \cdots j_l \cdots i_n}(k) - \widetilde D T_{i_1 \cdots i_n}(k)
  \cr
 \widetilde M^i_{~~ j} \equiv &  (  b^i \partial_{k^j} - b^j \partial_{k^i} )
\cr
\widetilde D \equiv &   (b . k ) {\vec \partial }_k^2  - 2  k_j \partial_{k_j} (b .  \partial_{k} )
\end{align}
Where the $(-3)$ in $\Delta -3$ comes from the commutators we had in \eqref{sch} .

In momentum space any generator has an overall momentum conserving delta function $\delta ( \sum_I \vec k_i)$. It is possible to pull the momentum space operator through the delta function. One can show
that all terms involving derivatives of the delta function vanish. This is argued in detail in
appendix \ref{DeltaFunction}. The final result is that we can simply act with the
operator \eqref{changep} on the coefficient of the delta function.

We would now like to express the action of the special conformal generator in terms of the spinor
helicity variables. This problem is very similar to the one analyzed for amplitudes in \cite{Witten:2003nn}.
There it was shown that the special conformal generator is given by
\begin{equation}\label{spac}
 b. \widehat{\mathcal{O}} \equiv b^i {\sigma^i} {}^{ a \dot a}  { \partial^2 \over \partial \lambda^a \partial \bar \lambda^{\dot a}}
\end{equation}
The closure of the algebra implies that this simple form can only be consistent when it is applied
to objects of scaling dimension minus one (in Fourier space).
In our case,
  we will see that \eqref{spac} differs from the special conformal generator only by
   terms proportional to  the Ward identity for the corresponding tensor (the current or the stress tensor). This will be discussed in more detail below.

\subsection{Constraints of special conformal invariance on scalar operators}

The correlation function of three scalar operators is very simple in position space and it is
given by a well known formula. In momentum space, it is hard to find an explicit expression
because it is hard to do the Fourier transform.
For the case of the three point function, the answer is a function of the $|\vec k_I|$.
In that case we can rewrite the special conformal generator as
\begin{equation} \label{specfn}
 \vec b . \vec k \left[ -2 ( \Delta -2 ) { 1 \over |k| } \partial_{|k|} + { \partial^2 \over
 \partial |k|^2 } \right]
 \end{equation}

 We see that the case of $\Delta =2$ is particularly simple\footnote{Note that in this case the Fourier
  transform has dimension minus one, and then the special conformal generator is given by the simple
  expression in \eqref{spac}.}. So we consider a situation with three scalar operators of dimension
  $\Delta =2$.
 Invariance under special conformal transformations then implies
\begin{equation}\label{original}
k_1^i \partial_{k_1}^2 f +k_2^i \partial_{k_2}^2 f+k_3^i \partial_{k_3}^2 f=0
\end{equation}
where $f$ is the Fourier transform of the correlator.
Using  momentum conservation   we can conclude that   all second
derivatives should be equal, for any $m\ne n$:
\begin{equation}\label{diffeq}
(\partial^2_{k_m}-\partial^2_{k_n}) f(k_1,k_2,k_3)=0
\end{equation}
For each pair of variables this looks like a two dimensional
 wave equation. Thus the general solution is given by
  $f(k_1, k_2, k_3) = g(k_1+k_2+k_3)+h(k_1-k_2-k_3)+l(k_2-k_3-k_1)+m(k_3-k_1-k_2)$. To fix the form of these functions we look at the dilatation constraint:
\begin{equation}\label{dilatationc}
(k_1\partial_{k_1}+k_2\partial_{k_2}+k_3\partial_{k_3})f(k_1,k_2,k_3) = c
\end{equation}
In principle, $c=0$. We would be tempted to conclude that this implies that each of the functions in $f$ should be scaling
invariant. This would leave only a constant solution. One can see that a logarithm is also allowed.
The variation of a logarithm is a constant, and in position space, this is just a contact term. In other
words, we can allow $ c \ne 0$ in the right hand side of \eqref{dilatationc}. Another way to see this is to consider the Fourier transform of the dilatation constraint. When we substitute $ x \to i \partial_k$ we are implicitly integrating by parts. In general we neglect the surface terms because they are not singular. In the case we are considering, one can see that these terms are non-zero, hence $c \ne0$.

In order to fix the combination of logarithms we can impose permutation symmetry as well as a
good OPE expansion. The OPE expansion in position space says that $ \la OOO \ra \sim { 1 \over x_{23}^2 }{ 1 \over x_{12}^4}$ as $x_{23} \to 0$. This translates into $ \la OOO \ra \sim
{ |\vec k_1| \over |\vec k_3| } $ as $\vec k_1 \to 0$  \footnote{ This OPE requirement is imposed up to
contact terms. Thus, for example, a term of the form $\log k_2$, in this limit gives us a $\delta^2(x_{12})$ (since it is independent of $k_1$)
and is consistent with the OPE requirement, which is only imposed at separated points.
In other words, when we expand \eqref{solution} for small $\vec k_1$ we get $ 2 \log k_2 + { k_1 \over 2 k_2 }  + \cdots $, we drop the first term and the second leads to the correct OPE. }.  We then find that
only the following solution  is allowed
\begin{equation}\label{solution}
\begin{split}
f(k_1,k_2,k_3)&  \sim \log
 (
 k_1 + k_2 + k_3 )
\end{split}
\end{equation}
It is possible to check that this is also the Fourier transform of the usual position space
expression, $ { 1 \over x_{12}^2 x_{13}^2 x^2_{23} }$.
It is also possible to show that one has simple solutions when operators of $\Delta = 2,1$ are involved.
This is done as follows. After we obtain \eqref{solution} we can express the Fourier transform of the
three point function as $f = \prod_{I=1}^3 k_I^{\Delta_I -2 } g $. Then $g$ has scaling dimension zero, and
the special conformal generator on each particle acquires the form
\begin{equation}
 \label{specfn}
\left(\prod_{I=1}^3 k_I^{\Delta_I -2 } \right) \vec b . \vec k \left[  {  - ( \Delta -2 ) ( \Delta -1) \over |k|^2 }  + { \partial^2 \over
 \partial |k|^2 } \right] g
 \end{equation}
We then see that for $\Delta =1,2$ the computation is the same as what we have done above. If all
operators have $\Delta=1$, then the answer is  $g=1$ or
$f_{\Delta =1} = { 1 \over k_1 k_2 k_3 } $.  When some operators have $\Delta =1$ and some $\Delta =2$
we cannot use permutation symmetry to select the solution, but it should be simple to find it.

\subsection{Constraints of special conformal invariance on conserved currents}

The constraints for special conformal invariance in momentum space are given by \eqref{changep}.
Here we would like to express the constraint of special conformal invariance in the
spinor helicity variables. We would like to express the special conformal generator in terms
of a simple operator such as \eqref{spac}. The operator we want to consider is the current in
Fourier space, multiplied by a polarization vector proportional to
$ \xi^-{}^{a \dot b } = { \lambda^a \lambda^{\dot b}\over
 \la \lambda , \bar \lambda \ra}$.
 In fact, just multiplying by this vector has a nice property, it leads to an operator of
 dimension minus one, since the Fourier transform of a conserved current has dimension minus one, and
 this choice of polarization vector does not modify the scaling dimension.
If $J$ is the conserved current, we take $\xi^- . J$ and we act with the operator \eqref{spac}.
The lambda
derivatives can act on $\xi$ and also on $J$, when they act on $J$, we can express them in terms
of $\vec k$ derivatives. After a somewhat lengthy calculation,
 one can rearrange all terms so that we get the action of \eqref{changep}
on the current, plus a term proportional to the divergence of $J$, or $\vec k . \vec J$.
More explicitly, we find\footnote{$\widetilde M^i_{~j} - \delta^i_j$ and $ \widetilde D $ are defined in
\eqref{changep}.}
\begin{equation}\label{currco}
  b^i {\sigma^i} {}^{ \beta \dot \alpha }  { \partial \over \partial \lambda^\beta } { \partial \over \partial \bar \lambda^{\dot \alpha} } ( \xi^- . J ) = \xi^-_i
  ( \delta^i_j 2 b.\vec \partial +
2 \widetilde M^i_{~j} - \delta^i_j \widetilde D ) J^j  - ( b.\xi^-) { k^i \over |\vec k|^2 } J^i
\end{equation}
The first term in the right hand side vanishes due to the special conformal generator. The second
term in the right hand side involves a longitudinal component of the current. One is tempted to set that
to zero. However, we should recall that, inside a correlation function, we get contact terms at the
positions of other charged operators.
 These terms are simply given by (the Fourier transform of) the Ward
identity
\begin{equation} \label{WIcu}
k^i_1 \la J^i(k_1) O_2(k_2) \cdots O_n(k_n) \ra =- \sum_{l=2}^n Q_l  \la O_2(k_2) \cdots O_l( k_l + k_1) \cdots O_n(k_n) \ra
\end{equation}
Where $Q_l$ is the charge of the operator $O_l$.
These are lower point functions. The conclusion is that there is a simple equation we can
write down, by acting with the special conformal generator in spinor helicity variables, \eqref{spac}.

Note that in position space, we normally impose the special conformal transformation at separated
points. In other words, we do not consider local terms. A local term will contribute to the three point function in position space as
\begin{equation}\label{localterm}
\la O(x) O(y) O(z) \ra_{Local} \sim {\cal D} [\delta^3(x-y)] f(x-z)+\mbox{cyclic}
\end{equation}
$ {\cal D}$ is an arbitrary differential operator, which, when integrated by parts, will just yield an analytic function of the $k$s (like $k_1^i$, $k_1^4 k_3^i$, ...). Upon Fourier transforming the first term, we see that its form will be $[\mbox{analytic piece}]\times F(k_3) + \mbox{cyclic}$, where $F$ is the Fourier transform of $f$. So, any piece in the three point function that is analytic in two of its variables, like $k_1$, $k_1k_2^2$, $k_3^4/k_1$, corresponds to a local term. Something like $k_1 k_2$ is analytic in $k_3$ but not in $k_1$ and $k_2$, so it is non-local.

Let us see how this works more explicitly.
We can start with the $--$ two point function
\begin{equation}\label{twopt}
\la \xi^- .J(k_1)  \xi^- . J(k_2) \ra = \delta^3(k_1 + k_2) { \la 1,2\ra^2 \over \la 1,\bar 1 \ra }
\end{equation}
Here the right hand side of the Ward identity vanishes and indeed, this
 function is  annihilated by \eqref{spac}.

Now we consider the three point functions of currents $J^a$ which are associated to a non-abelian
symmetry. In the bulk they   arise from a non-abelian gauge theory.
The Ward identity is given by
\begin{equation} \label{WIcur}
k_1^i\la J^a_i(k_1)J^b_j(k_2)J^c_l(k_3)\ra = f^{abc}[\la J_j(k_1+k_2) J_l(k_3) \ra - \la J_j(k_2) J_l(k_3+k_1) \ra]
\end{equation}
Where the color factor was   stripped off the two point correlators \footnote{One unpleasant
 feature of this equation, \eqref{WIcur},
is that the currents in the right hand side are evaluated at a shifted momentum, so it is not trivial
to express this in terms of the $\lambda$ and $\bar \lambda $ variables. In this particular case,
this is not a problem since the two point functions can be explicitly computed, but this might lead
to a more complicated story in the case of higher point functions.}.

Just as a check, let us compute
 the three point function for a gauge theory with a Yang Mills action in the
bulk. Since the gauge  field is conformally coupled, we can
do the computation in flat space.
We   compute this correlator  between three gauge fields in flat space, all set at $t=0$ and Fourier
transformed in the spatial directions.
 We have the usual    $f^{abc}$ non-abelian coupling in the bulk.
In Feynman gauge, the final answer is
\begin{equation}\label{rescor}
\la A^{a_1}_{\mu_1}(k_1)  A^{a_2}_{\mu_2}(k_2)  A^{a_3}_{\mu_3} (k_3) \ra \propto
\delta^3 ( \vec k_1 + \vec k_2 + \vec k_3 )
{ f^{a_1 a_2 a_3 } \over |k_1| |k_2| |k_3| }{ 1 \over E }
[  \delta^{\mu_1 \mu_2 } ( k_1^{\mu_3 } - k_2^{\mu_3} ) + {\rm cyclic } ]
\end{equation}
where $E =\sum_I |k_I| $.\footnote{
This is done as follows. The three point function in the bulk is the term in square brackets.
We attach the propagators. We write the energy conservation condition as
$\int dt' e^{ i t' \sum k^0_n }$.
Then we integrate over $k_n^0$ to localize things at $t=0$. We can deform the integration contour and
we only pick the residues of each of the poles of the propagators which give rise to the factor in
the denominator. We close the contour up or down depending on the sign of $t'$. In each case,
 integrating over
$t'$ gives us the factor of $1/E$. }

 We now multiply by $  \xi^-  $ for each particle
  to compute the $---$ correlator
  and we get
\begin{align}\label{rescorfin}
\la \xi_1^- . A^{a_1} (k_1)  \xi_2^- . A^{a_2} (k_2)  \xi_3^- . A^{a_3}  (k_3) \ra
& \propto \delta^3 ( \vec k_1 + k_2 + k_3 )
{ f^{a_1 a_2 a_3 } }{ \la 1,2 \ra\la 2,3 \ra\la 1,3 \ra  \over [\la 1, \bar 1 \ra \la 2, \bar 2 \ra \la 3,\bar 3 \ra  ]^2  }
 \end{align}
Note that the expectation value we started from, \eqref{rescor}, is not gauge invariant.
 On the other hand, once we put in the polarization vectors and we compute the transverse part, as in
 \eqref{rescorfin}, we get a gauge invariant result (under linearized gauge transformations).
Note now that this expectation value is related to the currents, via a formula similar to \eqref{threeTran}, which introduces extra factors of the two point function, which here are
simply factors of $|\vec k|$.
Thus we find
\be  \la  {\xi_1 } . J^{a_1} (k_1)  {\xi_2  } . J^{a_2} (k_2)  {\xi_3  } . J^{a_3}  (k_3) \ra
\propto
  k_1 k_2 k_3 \la \xi_1 . A^{a_1} (k_1)  \xi_2 . A^{a_2} (k_2)  \xi_3 . A^{a_3}  (k_3) \ra
 \label{curmom} \ee
 Now let us check that this expectation value obeys the conformal Ward identity, with the operator \eqref{spac}. The action of \eqref{spac} on the first current is
\begin{align}
b^i \sigma^{i a \dot a}\frac{\partial^2}{\partial \lambda_1^a \partial \bar \lambda_1^{\dot a}}\left[{ \la 1,2 \ra\la 2,3 \ra\la 1,3 \ra  \over   \la 1, \bar 1 \ra \la 2, \bar 2 \ra \la 3,\bar 3 \ra }\right]=& b^i \sigma^{i a \dot a}\left[\frac{\lambda_{2 a} \lambda_{1 \dot a} \la 2 , 3 \ra \la 3 , 1\ra }{\la 1, \bar 1 \ra^2 \la 2 ,\bar 2 \ra \la 3, \bar 3 \ra}-\frac{\lambda_{3 a} \lambda_{1 \dot a} \la 1,2 \ra \la 2 , 3\ra}{\la 1, \bar 1 \ra^2 \la 2, \bar 2 \ra \la 3, \bar 3 \ra}- \right. \nonumber \\& \left.-\frac{2 \bar \lambda_{1 a} \lambda_{1 \dot a}  \la 1, 2 \ra \la 2,3 \ra\la 3, 1 \ra}{\la 1, \bar 1 \ra^3 \la 2, \bar 2 \ra \la 3, \bar 3 \ra}\right]
\end{align}

We now use the Schouten identity - a consequence of the fact that the spinors live in a 2D space - to simplify this further. Expressing $\lambda_{2 a}$ in terms of $\lambda_{1 a}$ and $\bar \lambda_{1 a}$ we have $\la 1, \bar 1 \ra \lambda_{2 a} = - \la \bar 1, 2\ra \lambda_{1 a} + \la 1,2 \ra \bar \lambda_{1 a}$ and thus:
\begin{equation}
\frac{\lambda_{2 a} \lambda_{1 \dot a} \la 2 , 3\ra \la 3 , 1\ra}{\la 1, \bar 1 \ra^2 \la 2 ,\bar 2 \ra \la 3, \bar 3 \ra}=\frac{\bar \lambda_{1 a} \lambda_{1 \dot a}  \la 1, 2 \ra \la 2,3 \ra\la 3, 1 \ra}{\la 1, \bar 1 \ra^3 \la 2, \bar 2 \ra \la 3, \bar 3 \ra}-\frac{ \lambda_{1 a} \lambda_{1 \dot a} \la \bar 1, 2 \ra\la 2,3 \ra\la 3, 1 \ra}{\la 1, \bar 1 \ra^3 \la 2, \bar 2 \ra \la 3, \bar 3 \ra}
\end{equation}
We can do the same for $\lambda_{3 a}$ and then all that remains is a term proportional to $\lambda_{1 \dot a} \lambda_{1a}$, given by
\begin{align} \label{sigthr}
b^i \sigma^{i a \dot a}\frac{\partial^2}{\partial \lambda_1^a \partial \bar \lambda_1^{\dot a}}\left[{ \la 1,2 \ra\la 2,3 \ra\la 1,3 \ra  \over   \la 1, \bar 1 \ra \la 2, \bar 2 \ra \la 3,\bar 3 \ra  }\right]=& -\frac{b.\xi^-_1}{\la 1, \bar 1 \ra^2 \la 2, \bar 2 \ra \la 3, \bar 3 \ra}\left[\la \bar 1, 2\ra\la 2,3 \ra\la 3, 1 \ra-\la \bar 1, 3)\la 1,2 \ra\la 2,3 \ra\right]
\end{align}

Using the momentum conservation condition we can express ``cross-products" of the form $\la m, \bar n \ra$ for $m \ne n$ in terms of other brackets (the details are worked out in an appendix) through $(k_1+k_2+k_3)\la m, \bar n\ra= -2 \la m, o\ra\la \bar o, \bar n\ra$, where $m \ne n \ne o$ and then the term in \eqref{sigthr} is
\begin{align}
\la \bar  1, 2\ra \la 2,3 \ra\la 3, 1 \ra&-\la \bar 1, 3\ra\la 1,2 \ra \la 2,3 \ra=
- \la 2,3 \ra^2 (k_2 - k_3)
\end{align}
Putting the pieces together, this is the expected contribution from the Ward identity
\begin{equation} \begin{split}
\sum_{I=1}^3b^i\widehat{\mathcal{O}}_I^i& \left[\la \xi_1.J^a(k_1)  \xi_2.J^b(k_2) \xi_3.J^c(k_3) \ra \right] =
\\
= &  f^{abc}  \left\{  \frac{b.\xi_1}{\la 1, \bar 1 \ra^2}
{ \la 2,3 \ra^2 \over \la 2,\bar 2 \ra \la 3,\bar 3 \ra }  [\la 2, \bar 2 \ra - \la 3, \bar 3 \ra]+ {\mbox cyclic} \right\}
\\
= &  f^{abc}  \left\{  \frac{b.\xi_1}{k_1^2}
{\xi_2.\xi_3 }  [k_2 - k_3]+ {\mbox cyclic} \right\}
\end{split}
\end{equation}

 The expectation values that
 we have computed for gauge fields in de Sitter can also be computed in flat
 space, since gauge fields are conformal invariant\footnote{This is true for the tree correlators we are
 discussing but it is not true if loop corrections are taken into account.}. So, we are simply computing
 correlation function of  gauge invariant field strengths
  in $R^4$ but on a particular spatial slice. We are putting all
 operators at $t=0$. We can think in momentum space and consider the operators $F_{ab}(t=0,\vec k)$,
 $F_{\dot a  \dot b}(t=0,\vec k)$ where we Fourier transformed in the spatial coordinates but not
 in the time coordinate. Given $\vec k$ for each operator, we can define
  $\lambda$ and $\bar \lambda $ via \eqref{defllb}. We then can write the operators we considered above
  as:
  \be
  2 \xi^- .A = -{ 1 \over k }   \lambda^a  \lambda^{b} (
 F^+_{ab} + F^-_{\dot a \dot b} ) = -i \xi^-_i F_{jl} \epsilon^{ijl} ~,~~~~~~2 \xi^+ . A = { 1 \over k}  \bar \lambda^{\dot a}  \bar  \lambda^{\dot b} (
 F^+_{ab} + F^-_{\dot a \dot b} )  = i \xi^+_i F_{jl} \epsilon^{ijl}
 \nonumber \ee
  In both of these expressions,
  when we write $F_{ab}$,
  or $F_{\dot a \dot b}$ we mean the self dual and anti-self dual parts, but the indices are summed
   over with the indices of the indicated  $\lambda$'s.
   These expressions involve
  contractions that are not natural in flat space, but are reasonable once we break the full Lorentz
  symmetry to the rotation group.  These operators are set at $t=0$, and in Fourier space in the spatial
  section, with momentum $\vec k$.\footnote{ Note that both the self dual and anti-self dual parts of $F$ contribute
  to each of the helicities. The reason is that we have defined  a four momentum $(|k|,\vec k)$ in order to define $\lambda$, $\bar \lambda$. (We could also have reversed the sign of the zeroth
  component to $|k| \to -|k|$, which exchanges $\bar \lambda \leftrightarrow \lambda$). With this definition, the time component of this four momentum is not necessarily equal to the
  total four momentum of on shell waves coming in or out of the $F$ insertions
   (it can differ by a sign).}

One can write higher derivative operators that give rise to three point functions that are annihilated by the special conformal generator \eqref{spac}. These operators would be $a Tr[F^3]$ and $b Tr[\widetilde F F^2]$.  Now it is important to put in the $dS$ metric. The wavefunctions for $A$ are still the same as those in flat space, if we compute the three point functions perturbatively. These give the following three point functions\footnote{These couplings do not require a non-abelian theory. They are antisymmetric in the Lorentz indices, thus they require an antisymmetric
tensor. Thus, we could have three abelian
gauge fields $F^{I}$ and then the cubic couplings $ \epsilon_{IJL} tr[F^{I} F^J F^L]$, where the
trace is over the Lorentz indices.} 
 \begin{align} \label{fcubed}
 \la J^{a,+}(1) J^{b,+}(2) J^{c,+}(3) \ra & \propto (2\pi)^3 \delta^3(\sum k_i) (a+i~ b) f^{abc}{ \la \bar 1, \bar 2\ra \la \bar 2, \bar 3\ra \la \bar 3, \bar 1 \ra \over
(\la 1, \bar 1 \ra + \la 2,\bar 2\ra + \la 3, \bar 3 \ra)^3}\\
  \la  J^{a,-}(1) J^{b,-}(2) J^{c,-}(3) \ra & \propto (2\pi)^3 \delta^3(\sum k_i) (a-i~ b) f^{abc}{ \la 1, 2\ra \la 2, 3\ra \la 3, 1 \ra \over
 (\la 1, \bar 1 \ra + \la 2,\bar 2\ra + \la 3, \bar 3 \ra)^3}
 \end{align}

The result \eqref{rescorfin}, which comes from the usual Yang Mills term  can be converted
into a correlator of curents of the  form 
 \begin{align} 
\langle J ^{a,+}(1) J^{b,+}(2) J^{c,+}(3)\rangle &= 
\left. { \delta^3 \Psi[A] \over \delta A^{a,+}(1) \delta A^{b,+}(2) \delta A^{c,+}(3) } \right|_{A=0} 
\cr
&  \propto \delta^3( \vec k_1 + \vec k_2 + \vec k_3) f^{a_1 a_2 a_3}{ \la 1,2 \ra\la 2,3 \ra\la 1,3 \ra  \over \la 1, \bar 1 \ra \la 2, \bar 2 \ra \la 3,\bar 3 \ra    }
\end{align}
 is completely analytic  in momentum space, and could be viewed as
arising from a factor  in the wavefunction of the form $ \Psi \sim e^{ Tr[ A \wedge A \wedge A ] } $. However, we would need
a term in the wavefunction with the opposite sign to remove the $---$ correlator. Thus, although it
looks like a local term, it does not seem possible to remove both the $+++$ and the $---$ correlator
with the same factor.  On the other hand, \eqref{fcubed} is definitely non analytic in momentum, due
to the $1/E^3$ singularity. 

The current correlators are derivatives of the wavefunction. The expectation values of $A$ can be obtained
from them. In that case the parity violating $b$ term in \eqref{fcubed} drops out.  

\subsection{Constraints of special conformal invariance on the stress tensor}

In this section we consider the constraints of conformal invariance in momentum space for the
stress tensor.

We multiply the stress tensor by a convenient polarization tensor
$  \epsilon^-_{ij} T_{ij} (k) $, with
  $ \epsilon^-_{ij} = \xi^-_i \xi^-_j  $.
 In order to study the action of the special conformal generator it is convenient to
   define an operator containing an extra power of $k$ as
 \be
 \widehat {T}^- = {\epsilon^-_{ij} T_{ij}(k)\over k}  ~,~~~~~~~~~
{\epsilon^-_{ij} } \equiv  {\xi^- _i\xi^- _j  } = { \lambda^a \lambda^b \lambda^{\dot a}  \lambda^{\dot b} \over \la \lambda ,\bar \lambda\ra^2 }
 \ee
  The power of $k$ was chosen so that
 the special conformal generator has the simple expression given by
   \eqref{spac}.  The expression for $\epsilon_{ij}$ is that one that would give a naturally normalized
   tensor,  when we take the reality conditions into account.
    Again, this operator does not quite annihilate the correlator, but it produces
   a term involving the Ward identity in  the right hand side.
    Although more laborious in terms of manipulations, the general ideas are the same as in the current case so we will be more brief in the details of the conformal symmetry check.

We find that \eqref{spac} acts on the stress tensor as
\begin{equation}\label{actiontens}
b.\widehat{\mathcal{O}} \hat T^-=b.\widehat{\mathcal{O}}\left[{\epsilon^-_{ij} T_{ij} \over k}\right] =
{\epsilon^-_{ij}  \over k}[    4 \widetilde M^j{}_l  -  \delta^j{}_l \widetilde D ] T_{il} -3 {1 \over k^3} b_i \epsilon^-_{ij} k_l (T_{lj}+T_{jl})
\end{equation}
The first term is what we expect from \eqref{changep} for the stress tensor, and it vanishes. The
second term can be computed by using the Ward identity. Again, such terms are analytic in some of the
momenta. So if we disregard terms that are analytic in the momenta, we can drop also the term involving
the Ward identity.

Let us first ignore this Ward identity terms and compute homogeneous solutions of the equation.
Let us consider first a general $---$ three point function of the stress tensor. Such a general
three point function is given by
\be \label{ansmmm}
\la \hat T^- \hat T^- \hat T^- \ra =
\delta^3( \sum \vec k_I) [\la 1,2 \ra\la 2,3 \ra\la 3,1 \ra]^2 f(k_1,k_2,k_3)
\ee
With  $f$ a symmetric function of dimension minus six.
After some algebra, using Schouten identities, we get
\begin{align}\label{factmm}
\sum_n(\sigma^i)^{a \dot a}\frac{\partial^2}{\partial \lambda^a_n \partial \bar \lambda^{\dot a}_n}&\left[(\la 1,2 \ra\la 2,3 \ra\la 3,1 \ra)^2 f(k_1,k_2,k_3)\right] = -2\frac{\xi_1^i}{k_1}\la 1,2 \ra\la 2,3 \ra^3\la 3,1 \ra[k_3-k_2]\partial_{k_1}f + \nonumber \\ &+ [\la 1,2 \ra\la 2,3 \ra\la 3,1 \ra]^2 \left[\frac{4}{k_1} \partial_{k_1} f + \partial^2_{k_1} f \right] k_1^i + \mbox{cyclic}
\end{align}
Although the $\xi$s are linearly independent, the $k$s are not. A convenient way to rewrite \eqref{factmm} is to choose special conformal transformation parameters $b^i$ to project out a few components. Let us take $b^i \sim ( \lambda_2^a \lambda_3^{\dot b} + \lambda_3^a \lambda_2^{\dot b} )^i  $, for example. This combination was chosen so that the time component of $b$  is zero.
 We find
\begin{align}  \nonumber
(\lambda_2 \lambda_3) . \widehat{\mathcal{O}}[(---)f] &= \la 1,2 \ra^2\la 2,3 \ra^3\la 3,1 \ra^2\left \{4 (\partial_{k_2} - \partial_{k_3})f + k_3 (\partial^2_{k_1} - \partial^2_{k_3})f - k_2(\partial_{k_1}^2 - \partial_{k_2}^2)f \right \}
\\
\to ~~~~ &   0=4 (\partial_{k_2} - \partial_{k_3})f + k_3 (\partial^2_{k_1} - \partial^2_{k_3})f - k_2(\partial_{k_1}^2 - \partial_{k_2}^2)f  \label{constrmm}
\end{align}
It is straightforward to check that the gravity result we obtained from a $W^3$ in interaction contribution gets annihilated by this operator. Such a contribution is simply $f = ( k_1 + k_2 + k_3)^{-6}$.
 The Einstein contribution is not annihilated. It gives a nonzero answer that matches the
 expected answer from the Ward identity.
One can solve the equation \eqref{constrmm} and its two other cyclic cousins by brute force.
One finds the expected solution, mentioned above, plus a new solution that has the form
\be \label{secondsol}
f={ 1 \over  [(k_1+k_2-k_3 ) (k_2+k_3-k_1 )(k_3+k_1-k_2 )]^2 }
\ee
 This new solution does not have the right
limit when $\vec k_1 \to 0$. Namely, if we start with $T_{ij}(\vec k_1)$, when the momentum goes to zero we
do not expect any singular term when $\vec k \to 0$. In fact, an insertion of $T_{ij}(\vec k =0)$ corresponds
to a constant metric or a change of coordinates. So, in fact this limit has a precise form. On the other
hand if we look at this limit in \eqref{secondsol}, we find that $f \sim 1/k_1^4$, which is too singular
compared to the expected behavior.

Let us now turn our attention to the $--+$ correlator.
We first  write an ansatz  of the form
\be \label{mmpoth}
\la \hat T^- \hat T^- \hat T^+ \ra =
\delta^3( \sum \vec k_n) { [ \la 1,2 \ra  \la 2,\bar 3 \ra\la 1,\bar 3 \ra]^{2} }
 g(k_1,k_2,k_3)
\ee
where $g$ is a homogeneous function of degree six \footnote{ In analogy to flat space,
 one is tempted to write this in terms
of $[ { \la 1 , 2\ra^3 \over \la 2,3 \ra\la 1,3 \ra } ]^2 $. This is easy to do using the identities in appendix \ref{spinexp}, but
we found simpler expressions in terms of \eqref{mmpoth}. }.
We can now use a trick to get the homogeneous solutions of the special conformal generator.
  We note that if we exchange $\lambda \leftrightarrow \bar \lambda$, then the sign of $|k|$ is changed,
but $\vec k$ does not change. Now, the ansatz for \eqref{mmpoth} differs from \eqref{ansmmm} precisely
by such a change in the third particle.
Thus, the two solutions for $g$ in \eqref{mmpoth} are simply given by the two solutions for $f$
but with $k_3 \to - k_3$.
More explicitly, the two solutions are
\begin{align}
g =& { 1 \over (k_1 + k_2 - k_3)^6}
 \\
g= & { 1 \over  [(k_1+k_2+k_3 ) (k_2+k_3-k_1 )(k_3+k_1-k_2 )]^2 }
\end{align}
However, now both solutions are inconsistent with the small $\vec k_i$ limit.
The first solution has a problem when $\vec k_1 \to 0$ and the second when $\vec k_3 \to 0$.
Thus, both are discarded since these limits are too singular.
Even though this trick of exchanging $\lambda \leftrightarrow \bar \lambda$ was useful
for generating solutions of the homogeneous equation, the full results for the correlators
are {\it   not} given by such a simple exchange.

In conclusion, we have shown that there are no other solution of the conformal Ward identities beyond
the ones we have already considered.
It remains to be shown that the Einstein gravity answer obeys the conformal Ward identity.
One can check that the Einstein  gravity answer is not annihilated by the operator \eqref{spac}.
It gives a nonzero term.
This term is indeed what is expected from the Ward identity for the stress tensor, which is the
second term in \eqref{actiontens}. Of course, this
is expected since   Einstein gravity  has these
symmetries. The relevant expressions are left to appendix \ref{wardcheck}.

\section{Remarks on field theory correlators}\label{freeCFT}

In this section we compute the free field theory three point correlation function for
scalars and fermions. This is very similar to what was done in \cite{Osborn:1993cr} in position space.
Here we work in momentum space. We will compare these expressions to the gravity ones computed above.
The idea is that by considering a theory of a free scalar and a theory with a free fermion we
obtain two independent shapes for the three point function of the stress tensor. Since we have computed
the most general shapes above, this will serve as a check of our previous arguments. In addition,
the momentum space expressions for the correlators might be useful for further studies.
The two correlation functions that we obtain for scalars or fermions are parity conserving.
Our results indicate that there can be field theories that give rise to the parity breaking contribution.
Such field theories are not free, and it would be interesting to find the field theories that produce
such correlators \footnote{ There are bounds similar to the ones derived in \cite{Hofman:2009ug} for the
parity breaking and parity conserving coefficients that appear in a three point function. The
parity conserving bounds were  considered in \cite{Buchel:2009sk}. These bounds can be derived by
considering a thought experiment where we insert a stress tensor at the origin with some energy and then
we look at the angular dependence of the energy one point function, as measured by calorimeters placed
at infinity. The stress tensor at the origin has spin $\pm 2$ under the SO(2) rotation group of the spatial plane. The energy one point function as a function of the angle is $\langle {\cal E }(\theta) \rangle \sim e^{ i 4 \theta } + 1 + e^{ - i 4 \theta} $, with coefficients that depend on the parts of the stress  tensor three point functions  that
 we have characterized as coming from $(W^+)^3 $, $R$, $(W^-)^3$. }.
We will concentrate here on the free theory case\footnote{Similar calculations involving the trace of the stress tensor were considered in \cite{McFadden:2010vh}.}.

The computation is in principle straightforward, one simple has to compute a one loop diagram with
three stress tensor insertions. The only minor complication is the proliferation of indices. To compute the diagram itself one needs to use the standard Feynman parametrization of the loop integral. In particular, it is convenient to use the following Feynman parametrization:
\begin{equation}
{1 \over A B C} = \int_0^\infty { 2 ~d\alpha ~d\beta \over (A + \alpha B + \beta C)^3}
\end{equation}
  The final expression will have several contractions of polarization tensors with 3-momenta, e.g. $\epsilon^1_{ij} \epsilon^2_{jk} \epsilon^3_{ki}$, $ k^2_i k^2_j \epsilon^1_{ij} k^3_l \epsilon^2_{lm} k^2_n\epsilon^3_{nm}$, etc. Then one needs to use the expressions presented in the appendix to convert these to spinor brackets. The final answers have a simpler form than the ones with the polarization tensors.
  
We treat first the scalar case and then the fermion case.

\subsection{Three point correlators for a free scalar}

The stress-energy tensor for a real, canonically normalized scalar field is
\begin{equation}
T_{ij}(x)=\frac{3}{4}\partial_i\phi(x)\partial_j\phi(x)-\frac{1}{4}\phi\partial_i\partial_j\phi(x)-\frac{1}{8}\delta_{ij}\partial^2\phi^2(x)
\end{equation}
 The two point function is, up to the delta function:
 \be
 \la T^+ T^+ \ra_\phi = \frac{k^3}{256}
 \ee
 The three point functions are
 \begin{align}
\la& T^+(k_1) T^+(k_2)T^+(k_3) \ra_{\phi} = \left[-\frac{k_1^3+k_2^3+k_3^3}{64} + \frac{(k_1 k_2 k_3)^3}{2(k_1+k_2+k_3)^6}-\nonumber \right.\\& \left.-{(k_1+k_2+k_3)^2 \over 128}\left((k_1 + k_2 + k_3) - {(k_1 k_2 + k_1 k_3 + k_2 k_3)\over (k_1 + k_2 + k_3) }-{k_1 k_2 k_3\over (k_1 + k_2 + k_3) ^2}\right) \right]{(\la \bar 1, \bar 2 \ra\la \bar 2, \bar 3 \ra \la \bar 3, \bar 1 \ra)^2\over k_1^2k_2^2k_3^2}\\
\la& T^+(k_1) T^+(k_2)T^-(k_3) \ra_{\phi} = \left[-\frac{k_1^3+k_2^3+k_3^3}{64} -\nonumber \right.\\& \left.-{(k_1+k_2-k_3)^2 \over 128}\left((k_1 + k_2 + k_3) - {(k_1 k_2 + k_1 k_3 + k_2 k_3)\over (k_1 + k_2 + k_3) }-{k_1 k_2 k_3\over (k_1 + k_2 + k_3) ^2} \right)\right]{\left(\la \bar 1,\bar 2 \ra \la \bar 2, 3 \ra \la \bar 1,  3 \ra \right)^2\over k_1^2k_2^2k_3^2}
\end{align}

\subsection{Three point correlators for a free spinor}

The stress tensor for a complex Dirac spinor is given by
\begin{equation}
T_{ij}(x)=\frac{1}{4}( \bar\Psi\gamma_i\partial_j\Psi-\partial_j\bar\Psi\gamma_i\Psi)+(i\leftrightarrow j)
\end{equation}

The two point function is
\be
\la T^+ T^+ \ra_\psi = \frac{k^3}{128}
\ee
The three point function is given by:
\begin{align}
\la& T^+(k_1) T^+(k_2)T^+(k_3) \ra_{\psi} = \left[-\frac{k_1^3+k_2^3+k_3^3}{64} - \frac{(k_1 k_2 k_3)^3}{(k_1+k_2+k_3)^6}-\nonumber \right.\\& \left.-{(k_1+k_2+k_3)^2 \over 64}\left((k_1 + k_2 + k_3) - {(k_1 k_2 + k_1 k_3 + k_2 k_3)\over (k_1 + k_2 + k_3) }-{k_1 k_2 k_3\over (k_1 + k_2 + k_3) ^2} \right) \right]{(\la \bar 1, \bar 2 \ra\la \bar 2, \bar 3 \ra \la \bar 3, \bar 1 \ra)^2\over k_1^2k_2^2k_3^2}\\
\la& T^+(k_1) T^+(k_2)T^-(k_3) \ra_{\psi} = \left[-\frac{k_1^3+k_2^3+k_3^3}{64} +\nonumber \right.\\& \left.-{(k_1+k_2-k_3)^2 \over 64}\left((k_1 + k_2 + k_3) - {(k_1 k_2 + k_1 k_3 + k_2 k_3)\over (k_1 + k_2 + k_3) }-{k_1 k_2 k_3\over (k_1 + k_2 + k_3) ^2} \right)\right]{\left(\la \bar 1,\bar 2 \ra \la \bar 2, 3 \ra \la \bar 1,  3 \ra \right)^2\over k_1^2k_2^2k_3^2}
\end{align}

\subsection{Comparison with the gravity computation }

We see that these results contain the general shapes discussed in gravity, but they
also have an extra  term proportional to $\sum k_i^3 $.
 This is a contact term. Namely,
 it is  non-zero only when some operators are on top of each other.  In
 position space we get  a delta function of the relative displacement between
two of the insertions. These terms are easily recognized in momentum space because they
are analytic in two of the momenta.
These contact terms represent an ambiguity in the definition of the stress tensor. There is no ambiguity
in taking the first derivative with respect to the metric. However, these contact terms involve a second
derivative with respect to the metric. So if we define the metric as $g =e^{\gamma}$ and we take
derivatives with respect to $\gamma$ we are going to get one answer. If we took $g = 1 + \gamma'$ and
took derivatives with respect to $\gamma'$ we would get a different answer. In fact, we have the
same ambiguity in the gravity results if we define $\gamma'_{ij} = \gamma_{ij} + { 1 \over 2} \gamma_{il} \gamma_{lj}$. In that case the two results will differ precisely by such a term.
It is interesting to note
that the non-gaussian consistency condition discussed in \cite{Maldacena:2002vr} does depend on this precise definition
of the metric, since a constant $\gamma$ gives rise to different coordinate transformations depending
on how we defined $\gamma$.
 The one derived in \cite{Maldacena:2002vr}  holds when the metric is defined in terms of $g = e^\gamma $.

We can note that if we have $n_\phi$ scalars and $n_\psi$ dirac fermions,  then
we have to sum the two contributions to the three point functions that we have written above .
If $ n_\phi = 2 n_\psi$ , then we see that the term going like $1/E^6$ cancels. This is the contribution
that comes from a $W^3$ term in the bulk. This combination is also the one that appears in a
supersymmetric theory. In fact, in a supersymmetric theory the three point function of the
stress tensor does not have any free parameters\footnote{Of course, to have supersymmetry we need to consider an $AdS$, rather than a $dS$ bulk.}.
 It is the same as the one given by the pure Einstein
theory in the bulk, which does not contain any $1/E^6$ terms. This is related to the fact that in
four flat dimensions supersymmetry forces the $+++$ and $---$ amplitudes to vanish \cite{Grisaru:1977px}.

\section{Discussion}

In this paper we have computed the  possible shapes for non-gaussianity for gravitational
waves in the de Sitter approximation. Though three possible shapes are allowed by the
isometries, only two arise in de-Sitter expectation values. The parity violating shape contributes
with a pure phase to the wavefunction and it drops out from expectation values.
The two parity conserving  shapes were given in equations \eqref{hhhR}, \eqref{hhhW3dS}.
One of these shapes is  given by the Einstein theory. The other shape arise from higher
derivative terms. Under general principles the other contribution can be as big as the Einstein term
contribution. Of course, in such a case the derivative expansion is breaking down. However, the symmetries
allow us to compute the three point function despite this breakdown. This is expected for an inflationary
scenario where the string scale is close to the Hubble scale. This  requires a weak string coupling,
 so that we get a small value of $H^2/M_{Pl}^2 \sim g_s^2$.
One of these shapes is parity breaking. These three point functions of gravitational waves are expected to
be small, having an $f_{NL-gravity} = { \la \gamma \gamma \gamma  \ra \over \la \gamma \gamma \ra^2 } $ of order one.
In a more realistic inflationary scenario, which includes a slow rolling scalar field, then we expect
that these results give the answer to leading order in the slow roll expansion. It would be interesting
to classify the general leading corrections to the graviton three point function in a general inflationary
scenario. This can probably be done using the methods of \cite{Cheung:2007st,Cheung:2007sv,Chen:2006nt,Weinberg:2008hq}.
 Here by assuming exact de Sitter
symmetry we have managed to compute the correction to all orders in the derivative expansion.

We have presented the result in terms of the three point function for circularly polarized gravitational waves.
We used  a convenient spinor helicity description of the kinematics.
These spinor helicity formulas are somewhat similar to the ones describing flat space amplitudes.
It would be interesting to see if this formalism helps in computing higher order amplitudes
 in de Sitter space.
The problem of computing gravitational wave correlators in de Sitter is intimately related with the
corresponding problem in Anti-de Sitter. (The two are formally related by taking $R^2_{dS} \to - R^2_{AdS}$,
where $R$ are the corresponding radii of curvature). Thus, all that we have discussed here also
applies to the $AdS$ situation.  
The dS wavefunction is related to the $AdS$ partition function. In this case all three shapes can arise.
The parity violating shape arises from the $\int W^2 \widetilde W$ term in the action.  
These three point correlators, for Einstein gravity in $AdS$, were computed
in  \cite{Arutyunov:1999nw}.
It would be interesting to see if the spinor helicity formalism is useful for computing higher
point tree level  correlation functions.
 It is likely to be useful if one uses an on shell method like the one
proposed in \cite{Raju:2010by, Raju:2011mp} \footnote{Raju's proposal \cite{Raju:2010by, Raju:2011mp}
 is only for $D>4$ dimensions, but it might be possible
that something similar exists in $D=4$. }. In the spinor helicity formalism that we have introduced,
we have defined the ``time'' component of the momentum to be $|\vec k|$. The choice of sign here
was somewhat arbitrary. When we are in four dimensional flat space, there is a simple physical
interpretation for the results we get by analytically continuing to $ - |\vec k|$, as exchanging
an incoming into an outgoing particle. It would be interesting to understand better the
interpretation for this analytic continuation of the correlators we have been discussing. This
continuation was
  important in \cite{Raju:2010by, Raju:2011mp}.

These computations of three point functions in de Sitter or anti-de Sitter are intimately related
to the computation of stress tensor correlators in a three dimensional field theory. In fact, the symmetries
are the same in  both cases. Therefore the constraints of conformal symmetry are the same. The
physical requirements are also very similar. The only minor difference is whether we require the two
point function to be positive or not, etc. But in terms of possible shapes that are allowed the discussion
is identical. Thus, our results can also be viewed as giving the three point correlation functions
for a three dimensional field theory. The position space version of these three point functions was
discussed in \cite{Arutyunov:1999nw, Bastianelli:1999ab, Giombi:2009wh, Giombi:2010vg}.  For some three point functions the position space version is much simpler. On
the other hand, for the stress tensor, the position space correlator has many terms due to the
different ways of contracting the indices \cite{Osborn:1993cr}. The momentum space versions we have
written here are definitely shorter  than the position space ones. They are a bit convention dependent
due to the contact terms. Thus, they depend on precisely how we are defining the metric to non-linear orders.
Here we have made a definite choice.
An elegant and simple way to write correlation functions is to go to the embedding space
formalism \cite{Dirac:1936fq,Ferrara:1973yt,Weinberg:2010fx}.
 It is likely that one can obtain relatively simple expressions for the three point
correlators using that formalism. On the other hand, the momentum space formalism might be useful
for constructing conformal blocks, since, in momentum space, there is only one state propagating in
the intermediate channel.

 {\bf Acknowledgments}

We thank N. Arkani-Hamed, A. Bzowski, 
T. Dumitrescu, D. Hofman, D. McGady, H. Naves, L. Senatore, J. Trnka, X. Yin, E. Witten, M. Zaldarriaga
 and A. Zhiboedov for useful discussions. GLP was supported by the Department of State through a Fulbright Science and Technology Fellowship and through the US NSF under
Grant No. PHY-0756966, and thanks the Stanford Institute for Theoretical Physics for hospitality during part of this work. JM was supported in part by the U.S. Department of Energy under Grant \#DE-FG02-90ER40542.

\appendix
\section{Expression for the three point function in terms of an explicit choice for polarization tensors}

\begin{figure}[h]
\begin{center}
\includegraphics[width=.5\textwidth]{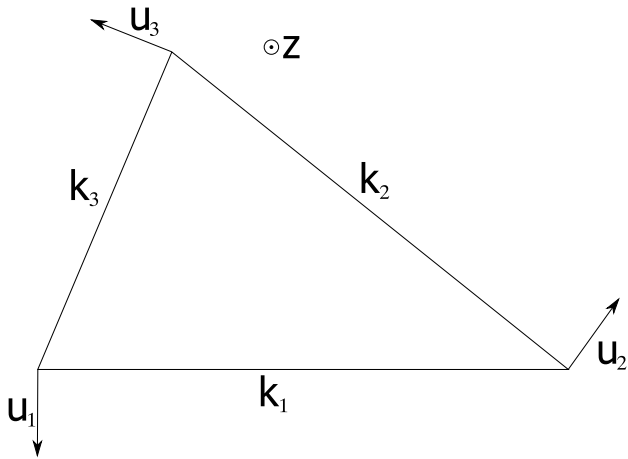}
\caption{The 3-momenta and the auxiliary vectors used to define the polarizations.}
\label{polarizations}
\end{center}
\end{figure}

As we are studying three point functions, there is a way to define polarization tensors that are similar to the usual ``$\times$" and ``$+$" of General relativity. We call them $X$ and $P$ here, so as not to cause confusion with the helicity labels $+$ and $-$. The choice of helicity states is based on the little group of an Euclidean 3D CFT. Basically, one takes two possible polarizations, P and X, as functions of a vector orthogonal to the plane of the triangle and of a vector orthogonal to one of the momenta we are looking at. So, using the notation defined in figure \ref{polarizations}, we have that
\begin{align}
\epsilon^{P,m}_{ij}&=2(z_iz_j-u^m_iu^m_j)\\
\epsilon^{X,m}_{ij}&=2(u^m_iz_j+z_iu^m_j)
\end{align}
And the previously discussed $+$ and $-$ polarizations will be given by $\pm=P\pm iX$. P and X are the polarizations known as $+$ and $\times$ in general relativity, but we choose to use different labels so that the former is not interpreted as positive helicity by mistake.

We list here the results for the non-gaussianities due to the Einstein term and the Weyl term.
The relevant pieces are labeled by the polarization choices $PPP$ and $XXP$. We always take particle three to have polarization $P$. The other structures are obtained by cyclic permutation. There are no $PPX$ and $XXX$ structures because they break parity, since $z$ flips under parity so that $X$ is odd and $P$ is even.
We use here the notation $J(k_1, k_2, k_3)\equiv 2(k_1^2k_2^2+k_1^2k_3^2+k_2^2k_3^2)-(k_1^4+k_2^4+k_3^4)$.

\begin{align}
&\la \gamma^{P}_{k_1}  \gamma^{P}_{k_2}  \gamma^{P}_{k_3} \ra_{R}= (2\pi)^3 \delta^3 \left(\sum_i k_i \right)\left(\frac{H}{M_{Pl}}\right)^4\frac{-1}{4(k_1k_2k_3)^5} \left[J(k_1,k_2,k_3)\left(\sum_{i=1}^3 k_i^4+6 \sum_{i < j}k_i^2 k_j^2\right)\right] \nonumber\\ &\left( k_1+k_2+k_3 - \frac{k_1k_2+k_1k_3+k_2k_3}{k_1+k_2+k_3}-\frac{k_1k_2k_3}{(k_1+k_2+k_3)^2}\right)\\ \nonumber \\
&\la \gamma^{X}_{k_1}  \gamma^{X}_{k_2}  \gamma^{P}_{k_3} \ra_{R}= (2\pi)^3 \delta^3 \left(\sum_i k_i \right) \left(\frac{H}{M_{Pl}}\right)^4\frac{1}{(k_1k_2k_3)^4} \left[J(k_1,k_2,k_3)\frac{k_1^2+k_2^2+3k_3^2}{k_3}\right] \nonumber\\ &\left( k_1+k_2+k_3 - \frac{k_1k_2+k_1k_3+k_2k_3}{k_1+k_2+k_3}-\frac{k_1k_2k_3}{(k_1+k_2+k_3)^2}\right)
\\ \nonumber \\
&\la \gamma^{P}_{k_1}  \gamma^{P}_{k_2}  \gamma^{P}_{k_3} \ra_{W^3}= (2\pi)^3\delta^3 \left(\sum_i k_i \right) \left(\frac{H}{M_{Pl}}\right)^6\left(\frac{H}{\Lambda}\right)^2 a\frac{2160}{(k_1+k_2+k_3)^4(k_1k_2k_3)^2}J(k_1,k_2,k_3)= \nonumber\\
&= (2\pi)^3\delta^3 \left(\sum_i k_i \right) \left(\frac{H}{M_{Pl}}\right)^6\left(\frac{H}{\Lambda}\right)^2 a\frac{270(k_1+k_2-k_3)(k_2+k_3-k_1)(k_3+k_1-k_2)}{(k_1+k_2+k_3)^3(k_1k_2k_3)^2}\\\nonumber\\
&\la \gamma^{X}_{k_1}  \gamma^{X}_{k_2}  \gamma^{P}_{k_3} \ra_{W^3}= - \la \gamma^{P}_{k_1}  \gamma^{P}_{k_2}  \gamma^{P}_{k_3} \ra_{W^3}
\end{align}

\section{Details on the spinor helicity formalism}\label{spinexp}

 Here we summarize some conventions that we have used.
\begin{itemize}
\item Metric: $ \eta_{\mu\nu}= \mbox{diag}(-1, +1, +1, +1)$; $\epsilon_{\dot a a}=\begin{pmatrix}0& -1 \\ 1 &0\end{pmatrix}$; $\epsilon^{a \dot a}=\begin{pmatrix}0& 1 \\ -1 &0\end{pmatrix}$

\item Sigma matrices: $\sigma^{\mu a}{}_b=(-\delta^a{}_b,\sigma^{ia}{}_b)$;

\item Scalar product: $p.q =-2 \la \lambda_p,\lambda_q \ra \la \bar\lambda_ p, \bar \lambda_ q \ra$; Energies: $p^0\equiv p =  - \la \lambda_p, \bar\lambda_ p  \ra = - \epsilon_{ab} \lambda_p^a \bar \lambda_p^b $

\item Polarization vectors used for the expressions in the appendix (normalization is not the same as the one used in the paper for the stress tensor): $\displaystyle \xi^-{}^{a \dot a }=-\frac{\lambda^{ a}\lambda^{\dot a}}{ k}$ and $\displaystyle \xi^+{}^{a \dot a} = \frac{\bar \lambda^a \bar \lambda^{\dot a} }{ k}$
\end{itemize}

Starting from a three momentum $\vec k$, we define a four momentum $k^\mu = ( |k|, \vec k)$.
This obeys $k^\mu k_\mu =0$. This defintion can be done for dS, AdS, or a three dimensional  CFT.
Note that $ k^i = \hat \sigma^{i}{}_{\dot a b } \lambda^b \bar \lambda^{\dot a} $, where $\hat \sigma^i$
are the Pauli matrices with an index lowered by $\epsilon_{\dot a c}$. These matrices are symmetric.
Thus we conclude that if we exchange $\lambda^a \leftrightarrow \bar \lambda^a$ we keep the
value of $\vec k$, but we change the sign of the energy $k^0 \equiv k$. This change is not consistent
with the reality conditions which are
$(\lambda^*)^a = \epsilon_{ \dot b a} \bar \lambda^{ \dot b}$, $(\bar \lambda^*)^{\dot b} =
- \epsilon_{\dot b a} \lambda^a$. However, if we just forget about the reality condition, then the exchange
of $\lambda \leftrightarrow \bar \lambda $ is allowed.
Note also that the following symmetry is consistent with the reality conditions and it reverses the
sign of $\vec k$ but does not change the sign of $k$. Namely, the exchange
$\lambda' = \bar \lambda$, $\bar \lambda' = - \lambda $. This is useful for the two point function,
where the momentum conservation condition forces the spatial momenta to be opposite.
Note that in some formulas we write $k_i$, which means $|\vec k|$. This can also be written in
terms of $k_i = - \la \lambda , \bar \lambda \ra $.

For a given 3-momentum $\vec{k}=(k_1,k_2,k_3)$ if we define $|\vec{k}|\equiv k_0$ then one can for example take the following explicit choice of spinors, assuming that the reality condition is satisfied:
\begin{equation}\label{expchoice}
\lambda^a = \left(\sqrt{{k_0+k_3 \over 2}}, {-k_1 + i k_2 \over \sqrt{2(k_0+k_3)}}\right)^T;~~~\bar \lambda^{\dot a}= \left({-k_1 - i k_2 \over \sqrt{2(k_0+k_3)}}, -\sqrt{{k_0+k_3 \over 2}}\right)
\end{equation}

Let us now summarize a few identities that are useful for the treatment of the three point function.
For the case of the three point function the 3-momentum conservation condition reads
\begin{equation}\label{3momc}
\lambda^a_1\bar \lambda^{\dot a}_1+\lambda^a_2\bar \lambda^{\dot a}_2+\lambda^a_3\bar \lambda^{\dot a}_3= - \frac{E}{2} \epsilon^{a \dot a} ~,~~~~~
E \equiv  k_1 + k_2 + k_3 = - \sum_{n=1}^3 \la \lambda_n , \bar \lambda_n \ra
\end{equation}
Where the coefficient two is determined from contracting the $a$ and $\dot a$. We can contract this expression with, say, $\lambda_1$ and $\bar \lambda_2$. The purpose of that is to derive an expression for an object of the form $(m ,\bar n)$, which has no interpretation as an energy, if $m\ne n$. By doing that we find that $2\la 1,3 \ra\la \bar 3 ,\bar 2 \ra = E(1, \bar 2 )$ and hence, we can write the general expression:

\begin{equation}
\la m ,\bar n\ra =- \sum_{o\ne m, n} \frac{2 \la m , o \ra \la \bar o, \bar n \ra}{ E}
\end{equation}

Where $m \ne n \ne o$. Also, note that the 4-momentum product of two distinct 4-momenta, in terms of the three energies and the total energy, is given by:
\be
 -2 \la m, n \ra \la \bar m, \bar n \ra =k^m_{~\mu}k^{n\mu}=-k^m k^n + {\bf k^m . k^n}=\frac{1}{2}(k_o-k_m-k_n)E
 \ee

And we also use the Schouten identity, which is useful to write a given spinor $\lambda$ in terms of two reference spinors $\mu$ and $\xi$
\be
\la \xi, \mu  \ra \lambda^a=\la \xi, \lambda \ra \mu^a - \la \mu, \lambda \ra \xi^a
\ee
Whenever we have an expression in terms of angle brackets we can write it completely in terms
of $\langle m, n \rangle$ brackets and the $k_m = - \langle m , \bar m \rangle $ brackets.
Useful identities to do so are (for $m \not = n$)
\be
\langle  m , \bar n \rangle  = {\langle  m,o\rangle \over \langle  n, o\rangle } {(k_o + k_n  - k_m) \over 2} ~,~~~~~~~~~ \langle \bar m , \bar n\rangle = -  {  (k_o - k_m - k_n) E \over 4\langle m,n \rangle  }
\ee

\subsection{Some expressions involving polarization vectors}
Now, let us calculate all the possible contractions of polarization vectors and momenta of different helicities:
\begin{align}
\xi^{+}_m.\xi^{+}_n&=-2\frac{\la \bar m, \bar n\ra^2}{ k_m k_n}\\
\xi^{-}_m.\xi^{-}_n&=-2\frac{\la m,  n \ra^2}{ k_m k_n}\\
\xi^{+}_m.\xi^{-}_n&=2 \frac{\la \bar m ,n\ra^2}{ k_m k_n}
\end{align}
As for contractions of momenta with polarization vectors, we have \footnote{As there is no time component of the polarization vector, $k^{a \dot a} \epsilon_{\dot a a} \sim k_{\mu}\epsilon^{\mu} = k_i\epsilon_i$ so we are only taking the space components into account.}
\begin{align}
k_m.\xi^{+}_n&=-\frac{2\la m, \bar n\ra \la \bar m, \bar n \ra}{  k_n} \\
k_m.\xi^{-}_n&=\frac{2 \la m , n \ra  \la \bar m, n \ra}{ k_n}
\end{align}

These can also be used to convert the gravity expressions into expressions in the spinor helicity
 variables since the gravity polarization tensor is $\epsilon_{ij} \sim \xi_i \xi_j$.

\section{Comments on the parity breaking piece of the two point function }
\label{ParityBreak}

As pointed out in \cite{Lue:1998mq} , the gravitational wave two point correlation function (or gravitational
wave spectrum) can be different for the two circularly polarized waves without breaking rotation
symmetry. In fact, a bulk coupling of the form $\int f(\phi) W \widetilde W$ is enough to produce this.
This mechanism requires an inflaton.
One can ask whether a parity breaking
two point function is possible in de Sitter space,
as some authors have suggested \cite{Contaldi:2008yz}.
Here we make some comments on the parity breaking pieces of the two point function of the stress
tensor.

The summary is that parity breaking terms are allowed in the gaussian part of the wavefunction of
the universe, or in the two point function of CFT's. However, such terms are local, and contribute
with a phase to the wavefunction. Thus they do not lead to different amplitudes for left and
right circular polarizations.

Let us start by discussing this from the wavefunction of the universe point of view. From that
point of view the question is  whether there can be a parity breaking two point function
for the stress tensor.  One is tempted to say that the answer is no. The argument is the following.
The stress tensor is in a
single representation of the conformal group, thus its two point function should be uniquely fixed.
In fact, this is correct if we consider the two point function at different spatial  points. However, there can
be a parity breaking contact term.
In order to understand this, let us discuss first the case of  a current, or a gauge field in the bulk, and
then discuss the case of the graviton or the stress tensor.

\subsection{Parity breaking terms in the two point functions for currents}

These were discussed in the $AdS$ context in \cite{Witten:2003ya}. We just summarize the
discussion here.
The Fourier transform of the conserved current two point function is (in a 3D CFT)
\begin{equation}\label{currentf}
 \la  J_i(k_1) J_j(k_2) \ra = \delta^3 ( k_1 + k_2) \left[  (\delta_{ij} k^2 - k_i k_j  ) |k|^{-1}  + \theta \epsilon_{ijl} k_{1,l}  \right]
\end{equation}
This is consistent with conformal symmetry. It is annihilated by
 \eqref{changep}. The $\theta$ term breaks parity. Since this term is analytic in the momentum, it
gives rise to a contact term in position space, a term proportional to $i \epsilon_{ijl} \partial_{x_l} \delta^3(x -y)$.
If we couple the current to an external source $A_\mu$ and compute $\Psi[ A] = Z[ A.J]$, then
we  are just adding a local term to the wavefunction of the Chern-Simons form
$\psi_\theta (A) = e^{  i \theta \int d^3x  A d A } \psi_{\theta =0} (A)$ \footnote{For the
non abelian case, we can complete this quadratic action into the full cubic one.}.
This is what we would get
in a $dS$ situation if we have an ordinary $\theta$ term in the bulk.  In other words, if we have a gauge
field in the bulk with an interaction $\theta \int Tr[ F \wedge F]$, then the wavefunction
contains a term   proportional to the Chern-Simons action on the spatial slice.
 In  a unitary (and gauge invariant) bulk theory this term has a real value of $\theta$, so that
it contributes as a phase in the wavefunction.
 Thus, when we compute the square of the wavefunction, this terms
drops out.
More explicitly,
 we can now compute the wavefunction in momentum space
 \begin{equation}\label{momspawf}{
 \psi(A) = \exp\{ - A_i ( k) A_j(-k)  (  \delta_{ij}  |k| - { k_i k_j \over |k| } +   \theta \epsilon_{ijs} k_s )  \}
 }
 \end{equation}
 We see that the $\theta$ term is imaginary if we take $\theta $ to be real. (We use that $A(k)^* = A(-k)$).
Then, if we compute $|\psi(A)|^2$ we  find that the $\theta$ term
 drops out if $\theta$ is
real.

Now we can ask, could it be that some unknown unitary Hamiltonian produces a
wavefunction that contains a Chern-Simons part
 with a purely imaginary $\theta$? The arguments leading to the Chern-Simons term in the wavefunction were based purely on demanding conformal symmetry and did not rely on any assumptions about
the bulk Hamiltonian, or even its existence. All we are assuming is that we have a wavefunction that
is conformal invariant.
In particular, purely from conformal symmetry, the $\theta$ term could be imaginary. An imaginary
$\theta$ leads to different amplitudes for the two circular polarization states of the gauge field\footnote{We are not assuming that we have an ordinary $\theta $ term in the bulk, but simply that
some unknown dynamics gives rise to a $\theta $ term in the wavefunction as in \eqref{momspawf}. }.

One problem with this is that
 the resulting probability amplitude is now not invariant under large
gauge transformations. This is due to the fact that the Chern-Simons action shifts by a certain
real factor
  under a large gauge transformation. Thus the wavefunctions produced in this way are not
gauge invariant
\footnote{This argument was suggested to us  by E. Witten.}. This argument is most clear in a non-abelian
situation.

However, if we ignore this problem, then we should also point out another issue.
A different amplitude for left and right circularly polarized waves violates CPT invariance\footnote{It is not known whether CPT invariance holds in quantum gravity. In
 theories that have a dual CFT description, like the ones in AdS, CPT is a good symmetry, so it seems reasonable to assume that CPT will still be a symmetry of dS.}.
This is most clear if we think about the observer in static patch coordinates. This observers sees de Sitter as static. For this observer   CPT is a symmetry, it is not spontaneously broken
by the background. However, CPT transforms $+$ circular polarization into $-$. Thus these amplitudes
cannot be different. Note that the wavefunctions that we discussed are the late times ones, the
wavefunctions for fluctuations outside the horizon. On the other hand, the static patch observer
probes the wavefunction inside the horizon. So, here we have assumed, by continuity, that if we get a
parity breaking effect outside the horizon, then we should also see some effect inside the horizon.

\subsection{Parity breaking two point functions for the stress tensor}

Similar arguments can be used for the two point function of the stress tensor in momentum space. The only term we can write that breaks parity, by power counting, is
\be \label{TTodd}
\la T_{ij}(k)T_{mn}(-k) \ra_{odd} \sim [(\epsilon_{iml} k_l \delta_{jn} k^2 + (i \leftrightarrow j))+(m \leftrightarrow n)]
\ee
 This is a function of $k_i$ and $k^2$, hence, it is analytic and corresponds to a local term in position space.

In gravity, an analog of the $\theta $ term is the topological invariant
\begin{equation}\label{topi}{
\int Tr[ R\wedge R] = \int \epsilon^{\mu \nu \rho \sigma} R^{a b}_{\mu \nu} R^{ab}_{\rho\sigma} =
\int   \epsilon^{\mu \nu \rho \sigma} R^{\gamma  \delta}_{~~~\mu \nu}
R^{\gamma \delta}_{~~~~\rho\sigma} = \int W \widetilde W
}
\end{equation}
The last equality follows from the symmetries of the Riemann curvature.
Adding this term as $\theta \int W \widetilde W $ to the bulk action we get
 a contribution to the wavefunction of the form
\begin{equation}\label{boundte}{
 e^{ i \theta S_{CS}(\omega) }
 }
 \end{equation}
 where $\omega$ is the spin connection. It can also be written in terms of the Christoffel connection
 \footnote{ In fact, we can use the relation between the two connections that comes from demanding that $D_\mu
  \epsilon^a_\nu =0$, which is
$
  \omega_\mu^{ab} = e^a_\alpha \Gamma^\alpha_{\mu \nu} E^{\nu b} - \partial_\mu e^a_\nu E^{\nu b}
 $
  In this form it has the form of a $GL(N)$ gauge transformation, $\omega_\mu = g \Gamma_\mu g^{-1} -
  \partial_\mu g g^{-1}$ with $g = e^a_{\alpha  }$. Of course, gauge transformations are a symmetry
  of the CS action. Thus we get the same action in terms of both connections. The first (upper) and last
  indices of $\Gamma$ are viewed as the ``internal'' GL(N) indices of the connection.}.
Let us check that this term indeed produces \eqref{TTodd}. We expand the Chern-Simons term to quadratic
order and obtain
 \begin{equation}\label{cstel}{
  S_{CS} \sim \int \epsilon^{ijk} \Gamma^{r}_{i s} \partial_k \Gamma^{s}_{j r}
  }
  \end{equation}
Using the first order expressions of the connection we find

\begin{equation}\label{actione}{
  S_{CS} \propto \epsilon_{ijl} ( \partial_r \gamma_{si} - \partial_s \gamma_{ri}) \partial_l\partial_r \gamma_{sj}
~~ ~~ \to ~~~~~
  \epsilon_{ijl} k^2 \gamma(k)_{si} k_l \gamma(-k)_{si}
  }
  \end{equation}
 where we used that $\gamma$ is transverse,
 $k_s\gamma_{si} =0$. This indeed reproduces \eqref{TTodd}.

As in the gauge field case, if $\theta$ is real, this term disappears from $|\Psi|^2$. On the other hand,
if $\theta$ is imaginary, we do get an extra contribution to $|\Psi|^2$ which leads to
a different amplitude for left and right circularly polarized gravitational waves, as pointed out in
\cite{Contaldi:2008yz}.

 Note that $S_{CS}$ in \eqref{boundte} is
 not invariant under large gauge transformations of the local Lorentz indices of the spin connection.

 Again CPT invariance forbids a different amplitude for left and right circular polarization.

Note that all the remarks in this section apply to the case of pure de Sitter. In the case that we
have an inflationary background, time reversal symmetry is broken by the inflaton and we can certainly
have a parity violating two point function \cite{Lue:1998mq}.

\section{Commuting through the delta function}
\label{DeltaFunction}

In this appendix we show that the action of the special conformal generator \eqref{changep} on a correlator or expectation value of the form $\delta(P) \mathcal{M}$ is equal to the action of the operator on $\mathcal{M}$. In other words, \eqref{changep} commutes with the momentum conserving delta function. The point is to understand how to get all the momentum derivatives through the momentum conserving
$\delta$ function. From now on all derivatives will be $k$ derivatives.

The delta function depends only on the sum of all momenta, let us call that $P$.
Then the sum over all particles of $D_a$ where $a$ runs over particle number has the form
\begin{equation}\label{deltaf}
\left(\sum_a \widetilde D_a \delta^3(P)\right) \mathcal{M}  = 6 [ (b . \partial_{\vec P } ) \delta^3 ] \mathcal{M}
\end{equation}
In order to derive this we have done the following. In each term the derivatives are with respect to
$k_a$, which end up $\partial_{P}$ when acting on the $\delta$ function. Then the $k_a$ in
$\widetilde D$ all sum up to $P$. Thus we have a term of the form $P ( \partial_P \partial_P \delta ) \mathcal{M}$.
We then integrate by parts the derivatives to act on ${\cal M}$ in such a way that we get terms of
the form in \eqref{deltaf} and also terms of the form $P \delta(P) \partial_P \partial_P \mathcal{M} $. Such terms
vanish.
Thus \eqref{deltaf} is the total contribution from terms with two derivatives on the delta function.

We can now consider terms which have only one derivative on the delta functions. There are terms
coming from $\widetilde D$. Let us consider those first.
The first term in $\widetilde D$ contributes with
\begin{equation}\label{first}
2 [ \partial_{P_j} \delta ] ( 2 b^i . k^i_a ) \partial_{k^j_a }
\end{equation}
The second term gives
\begin{equation}\label{secter}
2 [ \partial_{P_j} \delta ]\left\{ ( - 2 k_a^j  )( b . \partial_{k_a }  ) - 2 b^j ( k. \partial_k )
\right\}
\end{equation}
The \eqref{first} and the first term in \eqref{secter} give the action of the rotation generators on the term
multiplying the $\delta $ function. These would make the correlator vanish if it was rotational invariant.
On the other hand, we have indices, thus the correlator is rotational covariant. The action of the
rotation generators has to results in some action on the indices.  These must arise from the action of
$\widetilde M$ on the $\delta$ function, producing the spin generators.
Finally we have the last term, which adds up to the dilatation generator.
Such terms combine with a derivative of a $\delta $ function in \eqref{deltaf} and also a derivative that
comes from the first term in \eqref{changep} .
They altogether sum up to
\begin{equation}\label{sumallto}
 -( b . \partial_{P} \delta )  2 \left\{  \sum_a ( \Delta_a -3 )  ~~ + 3  - \sum_a k_a . \partial_{k_a}
\right\}
\end{equation}
The last term is computing the overall dimension of the term that multiplies the $\delta $ function,
the $+3$ is taking into account the dimensions of the $\delta$ function. And the first term is the
total dimension of the (Fourier transform) of the external states. Thus, if the answer is dilatation
covariant these terms will vanish. The total dilatation eigenvalue $ k \partial_k$ of the
coefficient of the delta function is then $3 + \sum_a ( \Delta_a -3)  $.

\section{Checking the Ward identities for the stress tensor}\label{wardcheck}

The Ward identities come from the statement that the wavefunction of the universe is
reparametrization invariant $\Psi[ g_{ij} ] = \Psi[ g_{ij } + \nabla_{(i }v_{j )} ]$.
This then implies that
\be \label{genWard}
 \nabla^{i} \left[ { 1 \over \sqrt{g } } { \delta \Psi \over \delta g^{ij} } \right]=0
 \ee
The stress tensor correlators are defined by taking multiple derivatives of the wavefunction and then
setting $g$ to the flat metric. By taking multiple derivatives of \eqref{genWard} we get
the Ward identity which looks like
\be
\partial_i \langle T_{ij}(x)  T_{l_1 ,m_1}(y_1) \cdots T_{l_n, m_n }(y_r) \rangle
 = \sum_{s=1}^r D^{l'_s m'_s}_{l_s m_s}  \delta^3(x-y_s)  \langle  T_{l_1 ,m_1}(y_1) \cdots T_{l'_s ,m'_s}(y_s)\cdots T_{l_n, m_n }(y_r) \rangle
\ee
where $D$ is a first order derivative (acting on $x$ or $y_s$)
and the indices are contracted in some way. These terms come
from acting with the metric derivatives on the explicit metric dependence in the covariant derivative, etc. Notice that if all the points in the left hand side are different from each other, then the right
hand side is zero. In this section we will assume that all the points $y_1, \cdots, y_r$ are different
from each other.
The precise form of the contact terms in the right hand side depends on the precise definition of
the ``derivative with respect to the metric''. If we define the stress tensor as derivatives
of the from ${ \delta \over \delta g^{ij}} $, then the Ward identities can be found in
\cite{Osborn:1993cr}.  However, in our case, we defined the stress tensor correlators as derivatives
of the wavefunction with respect to $\gamma_{ij}$, where we write the metric as $g = e^\gamma$, see
\eqref{WFU}. This leads to slightly different expressions. The difference is only present as extra
contact terms that arise when we use the chain rule $ T_{ij}^{\rm ours} =
{ \delta \over \delta \gamma^{ij}} = { \delta g^{lm}\over \delta \gamma^{ij}}{ \delta \over \delta g^{lm }}
$.

One can keep track of these extra terms and write the precise version of the Ward identity.

For the case of the three point function, after going to Fourier space, we get the simple expression
\footnote{This gets simplified thanks to the fact that $\la T_{ij}(-k) [\xi \xi T(k)] \ra = 2 \xi^i \xi^j k^3$.}
\begin{equation}\label{wardten}
\la [k_1 \xi^1
T(k_1)][\xi_2\xi_2 T(k_2)][\xi_3\xi_3 T(k_3)] \ra =   \xi_2.\xi_3\left\{\xi_1.k_3 \xi_2.\xi_3+\xi_2.k_1 \xi_1.\xi_3+\xi_3.k_2\xi_1.\xi_2 \right\} \left[2 k_2^3-2 k_3^3 \right]
\end{equation}
Which, for $---$ is given by:
\begin{equation}\label{wardfinalm}
\la [k_1 \xi^1
T(k_1)][\xi_2\xi_2 T(k_2)][\xi_3\xi_3 T(k_3)] \ra = 8 \la 1,2 \ra \la 2,3\ra^3\la 3,1\ra{(k_1+k_2+k_3)(k_2^3-k_3^3)\over (k_1 k_2 k_3)^2}
\end{equation}

From the point of view of the operator \eqref{spac}, this expression \eqref{wardfinalm} should be multiplying a term that goes like $(k_1^{-3}) \xi^1 (...)$. The coefficient of the three point function is then fixed by comparing this with the result of acting on the three point functions of Einstein gravity with \eqref{spac}.

For a general $---$ three point function, given by $(\la 1,2\ra \la 2,3 \ra \la 3,1\ra)^2 f(k_1,k_2,k_3)$, the action of \eqref{spac}  is \footnote{When there is no index in the derivative, it is understood that we are taking the derivative with respect to the energy, or $|k|$. We hope this does not cause confusion in the expressions here.}
\begin{align}
(\sigma^i)^{a \dot a}\frac{\partial^2}{\partial \lambda^a_1 \partial \bar \lambda^{\dot a}_1}&\left[(\la 1,2\ra \la2,3\ra \la3,1\ra)^2 f(k_1,k_2,k_3)\right] =\nonumber \\ &=[\la 1,2\ra \la2,3\ra \la3,1\ra]^2\left[(6 + 2 k_1^m \partial_{k_1^m}) \partial_{k_1^i} f -k_1^i \partial_{k_1^l}\partial_{k_1^l}f \right] +\nonumber\\&+i \epsilon^{ijk}\frac{\xi_1^k k_1^j}{k_1} [2 \la 1,2\ra \la 2,3\ra^2\la 3,1\ra[\la 2,\bar 1\ra \la 3,1\ra-\la 3,\bar 1\ra \la 1,2\ra ]]  \partial_{k_1} f
\end{align}
In order to derive this, we used Schouten to express $\lambda^2_a, \lambda^3_a$ each as a function of $\lambda^1_a$ and $\bar \lambda^1_a$.    As $\xi^1$ is a complex vector, one can show that $ i \epsilon^{ijk} \xi_1^k { k_1^j \over k_1} = \xi_1^i$. Then the third line is proportional to   $\xi^1$. The next step is to write the first piece as a function of $k_1$ and not of its components. The final result is:
\begin{align}\label{factmnp}
\sum_n(\sigma^i)^{a \dot a}\frac{\partial^2}{\partial \lambda^a_n \partial \bar \lambda^{\dot a}_n}&\left[(\la 1,2\ra \la2,3\ra \la3,1\ra)^2 f(k_1,k_2,k_3)\right] = -2\xi_1^i(1,2)(2,3)^3(3,1)[k_3-k_2]\partial_{k_1}f + \nonumber \\ &+ [\la 1,2\ra \la2,3\ra \la3,1\ra]^2 \left[\frac{4}{k_1} \partial_{k_1} f + \partial^2_{k_1} f \right] k_1^i + \mbox{cyclic}
\end{align}
Although the $\xi$s are linearly independent, the $k$s are not. A convenient way to rewrite \eqref{factmnp} is to choose special conformal transformation parameters $b^i$ to project out a few components. Let us take $b^i \sim (\lambda_2 \lambda_3+ \lambda_3 \lambda_2)^i$ for example. The constraint is \begin{align}\label{constraintmm}
(\lambda_2 \lambda_3) . \widehat{\mathcal{O}}[(---)f] =\la 1,2\ra^2\la2,3\ra^3\la3,1\ra^2\left \{4 (\partial_{k_2} - \partial_{k_3})f + k_3 (\partial^2_{k_1} - \partial^2_{k_3})f - k_2(\partial_{k_1}^2 - \partial_{k_2}^2)f \right \}
\end{align}
For the $--+$ correlator, the derivation is similar.

\bibliographystyle{hunsrt}
\bibliography{cosmonew}
\end{document}